\documentclass[12pt,preprint]{emulateapj}

\newcommand\jcap{JCAP}
\newcommand{\vrad}{$v_R$}

\newcommand{\logl}{log($L/{L_{\odot}}$)}

\newcommand{\teff}{$T_{\rm eff}$}

\newcommand{\pmra}{$\mu_{\alpha *}$}
\newcommand{\pmdec}{$\mu_{\delta}$}
\newcommand{\kms}{km\,s$^{-1}$}
\newcommand{\masyr}{mas\,yr$^{-1}$}

\newcommand{\mjup}{$M_{Jup}$}
\newcommand{\msun}{$M_{\odot}$}
\newcommand{\rsun}{$R_{\odot}$}

\newcommand{\logg}{log($g$)}

\newcommand{\vsini}{$v$sin$i$}

\newcommand{\And}{And}

\begin{document}

\title{The $\kappa$ Andromedae system: 
New constraints on the companion mass, system age \& further multiplicity}

\author{Sasha Hinkley\altaffilmark{1,17},
Laurent Pueyo\altaffilmark{2,18},
Jacqueline K. Faherty\altaffilmark{3},
Ben R. Oppenheimer\altaffilmark{4}, 
Eric E. Mamajek\altaffilmark{5},
Adam L. Kraus\altaffilmark{6},
Emily Rice\altaffilmark{7,3},
Michael J. Ireland \altaffilmark{8,9},
Trevor David\altaffilmark{1},
Lynne A. Hillenbrand\altaffilmark{1},
Gautam Vasisht\altaffilmark{10},
Eric Cady\altaffilmark{10},
Douglas Brenner\altaffilmark{4},
Aaron Veicht\altaffilmark{4},
Ricky Nilsson\altaffilmark{4},
Neil Zimmerman\altaffilmark{11},
Ian R. Parry\altaffilmark{12},
Charles Beichman\altaffilmark{13},
Richard Dekany\altaffilmark{14},
Jennifer E. Roberts\altaffilmark{10},
Lewis C Roberts Jr.\altaffilmark{10},
Christoph Baranec\altaffilmark{14},
Justin R. Crepp\altaffilmark{15},
Rick Burruss\altaffilmark{10},
J. Kent Wallace\altaffilmark{10},
David King\altaffilmark{12},
Chengxing Zhai\altaffilmark{10},
Thomas Lockhart\altaffilmark{10},
Michael Shao\altaffilmark{10},
R\'emi Soummer\altaffilmark{2},
Anand Sivaramakrishnan\altaffilmark{2},
Louis A. Wilson\altaffilmark{16}}
\altaffiltext{1}{Department of Astronomy, Caltech, 1200 E. California Blvd, MC 249-17, Pasadena, CA 91125}
\altaffiltext{2}{STScI, 3700 San Martin Drive, Baltimore, MD 21218, USA}
\altaffiltext{3}{Department of Astronomy, Universidad de Chile Cerro Calan, Las Condes, Chile, USA}
\altaffiltext{4}{Astrophysics Department, AMNH, Central Park West at 79th Street, New York, NY 10024, USA}
\altaffiltext{5}{Department of Physics and Astronomy, University of Rochester, Rochester, NY 14627-0171, USA}
\altaffiltext{6}{Harvard-Smithsonian CfA, 60 Garden St, Cambridge, MA 02140, USA}
\altaffiltext{7}{College of Staten Island, CUNY, 2800 Victory Bvld, Staten Island, NY 10314, USA}
\altaffiltext{8}{Department of Physics and Astronomy, Macquarie University, NSW 2109, Australia}
\altaffiltext{9}{AAO, PO Box 296, Epping NSW 1710, Australia}
\altaffiltext{10}{JPL, 4800 Oak Grove Dr., Pasadena CA 91109, USA}
\altaffiltext{11}{MPIA, Kšnigstuhl 17, 69117 Heidelberg, Germany}
\altaffiltext{12} {Institute of Astronomy, Cambridge CB3 0HA, UK}
\altaffiltext{13}{NExScI, California Institute of Technology, Pasadena, CA 91125}
\altaffiltext{14}{Caltech Optical Observatories, Pasadena, CA 91125, USA}
\altaffiltext{15}{University of Notre Dame, Dept. of Physics, 225 Nieuwland Science Hall, Notre Dame, IN 46556, USA}
\altaffiltext{16} {Washington University, St. Louis, MO, 63130}
\altaffiltext{17}{NSF Fellow, email: shinkley@astro.caltech.edu}
\altaffiltext{18}{Sagan Fellow}


\begin{abstract}
  $\kappa$ Andromedae is a B9IVn star at 52 pc for which a faint
  substellar companion separated by 55$\pm$2 AU was recently
  announced.  In this work, we present the first spectrum of the
  companion, ``$\kappa$ \And\, B,'' using the Project 1640
  high-contrast imaging platform.  Comparison of our low-resolution
  $YJH$-band spectra to empirical brown dwarf spectra suggests an
  early-L spectral type.
  Fitting synthetic spectra from PHOENIX model atmospheres to our
  observed spectrum allows us to constrain the effective temperature
  to $\sim$2000~K, as well as place constraints on the companion surface gravity. 
  Further, we use previously reported \logg\ and \teff\ measurements of the host star to argue that the $\kappa$ And system has an isochronal age of 220$\pm$100 Myr, older than the 30 Myr age reported previously.  
  This interpretation of an older age is corroborated by the photometric
  properties of $\kappa$ \And\, B, which appear to be marginally
  inconsistent with other 10-100 Myr low-gravity L-dwarfs for the
  spectral type range we derive.
  In addition, we use Keck aperture masking interferometry combined with
  published radial velocity measurements to rule out the existence of
  any tight stellar companions to $\kappa$ And A that might be responsible for the system's overluminosity. 
 Further, we show that luminosity enhancements due to a nearly ``pole-on'' viewing angle coupled with extremely rapid rotation is unlikely.
  $\kappa$ And A is thus consistent with its slightly
  evolved luminosity class (IV) and we propose here that $\kappa$
  \And, with a revised age of 220\,$\pm$\,100 Myr, is an
  interloper to the 30 Myr Columba association with which it was
  previously associated.  
  The photometric and spectroscopic evidence for
  $\kappa$ And B combined with our re-assesment of the system age
  implies a substellar companion mass of 50$^{+16}_{-13}$ \mjup,
  consistent with a brown dwarf rather than a planetary mass
  companion.
\end{abstract}


\section{Introduction}
Recent observations of young stars in the solar neighborhood,
employing high-contrast imaging techniques \citep[e.g.][]{am10,oh09}
have begun to determine the frequency and orbital distributions of
substellar and planetary-mass companions to nearby stars
\citep{mh09,nc10,lsh10,vpm12}. Observing the youngest systems, in
which substellar companions are still self-luminous during their
initial contraction, reduces the still formidable challenge of
overcoming the large brightness difference between the companion and
the host star.  Indeed, high-contrast observations in very young
($\sim$2-10 Myr) star forming regions have uncovered a handful of
wide-separation planetary-mass companions \citep[e.g.][Kraus et
al. {\it submitted}]{cld04, ljv08,ikm11}, although debate continues
regarding the exact nature of these objects.


\begin{figure}
\begin{center}
\epsscale{1.0}
\plotone{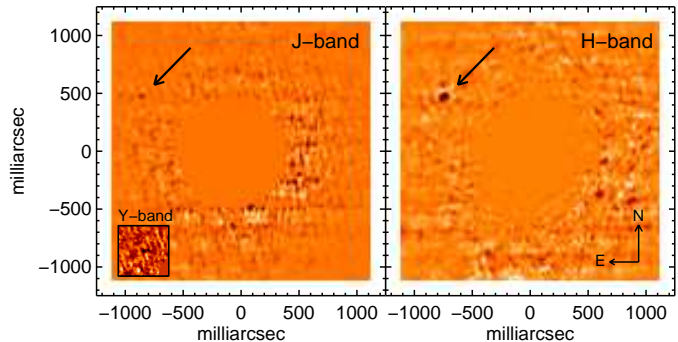}
\end{center}
\caption{A post-processed image obtained on 23 December 2012 from the
  Project 1640 high contrast imaging platform showing the $\kappa$ And
  B companion at the upper left.  
  }
\label{fig:P1640_s4}
\end{figure}

Further, some high contrast imaging surveys \citep[e.g.][]{vpm12,
  obb12,rcl13}
have been targeting nearby field and moving group stars. 
Assigning ages for
intermediate-mass, early-type stars is particularly challenging given
the relative immaturity of this field compared to solar type stars for
which many empirical age proxies are available. One such young,
intermediate-mass field star, Kappa Andromedae (hereafter, ``$\kappa$
And'') is a B9IVn star located at 52 pc for which a planetary-mass
companion, ``$\kappa$ And B'', was announced by
\citet{Carson13}. \citet{zrs11} claim that $\kappa$ And is a member of
the 30 Myr Columba association and using this assumption,
\citet{Carson13} derive a mass of 12-13 \mjup~for the companion.\
\footnote[1]{It is worth noting that our choice to use ``$\kappa$ And
  B'' to refer to the companion should not be confused with the
  purported {\it stellar} companions ``$\kappa$ And B'' and ``$\kappa$
  And C'' identified by \citet{h31}.  We note that the Washington
  Double Star (WDS) catalog refers to the companion reported in
  \citet{Carson13} as ``$\kappa$ And Ab'', since it is the fourth
  component of the $\kappa$ And ABC system to be discovered. However,
  as we describe in the appendix, it is exceedingly unlikely that the
  {\it stellar} components ``$\kappa$ And B'' and ``C'' identified by
  \citet{h31} are physical components of the $\kappa$ And
  system. Nonetheless, they are listed as such in the WDS. We choose
  to use ``$\kappa$ And B'' instead of ``Ab'' to remain consistent
  with \citet{Carson13}.  }


We begin with a discussion of the companion
(\S\ref{sec:secondary_props}), including a presentation of the first
spectrum of this object (\S\ref{sec:p1640_spectra}).  Comparing our
spectrum with empirical spectra of brown dwarfs
(\S\ref{sec:BDcomparison}) indicates the spectrum of this object is
consistent with spectra for an ``intermediate age'' ($\lesssim$300
Myr) low-gravity L1 brown dwarf, but similiarities with slightly later
spectral type ($\sim$L4) field objects remain. In
\S\ref{sec:synthetic}, we compare our data with synthetic model
spectra of substellar objects to constrain the surface gravity and derive a best-fit
\teff$\sim$2000K. 
Section~\ref{sec:nir_photometry}
presents our analysis of the near-infrared photometry of $\kappa$ And
B comparing its published $(J-K_s)$ color with the near-infrared
colors for low-gravity $\gamma$ L-dwarfs for the early-L spectral type
we derive.  In \S\ref{sec:props}, we review the properties of the host
star, $\kappa$ And A.  Using previously published \logg\ and \teff\
data for $\kappa$ And A, we use isochronal analysis to present a
revised system age (\S\ref{sec:age}). In addition, we show that
$\kappa$ And A is overluminous for a star with the originally assumed
age of 30 Myr, suggesting substantial evolution away from the zero-age
main sequence.  Using aperture masking interferometry
(\S\ref{sec:multiplicity}), we place stringent limits on the presence
of any stellar multiplicity that would be responsible for the
overluminosity.  We also show in \S\ref{sec:inclination} that a ``pole-on'' viewing angle, coupled with extremely rapid rotation, is unlikely for $\kappa$ And, which could also be responsible for the overluminosity.   Finally, given the disparity between our 220 Myr derived age
and the young 30 Myr age of the Columba Association with which it was
previously associated, in \S\ref{sec:kinematics} we re-examine the
kinematics of the $\kappa$ And system and suggest that the $\kappa$
And may in fact be an interloper to Columba.  Synthesizing this
information, our re-assessment of the key parameters of this system
implies a mass of 50$^{+16}_{-13}$ \mjup~for $\kappa$ And B,
consistent with a brown dwarf rather than a planetary mass companion.

\section{Properties of the Secondary: $\kappa$ And B}\label{sec:secondary_props}


\begin{figure*}[!ht]
\begin{center}
\epsscale{1.0}
\plotone{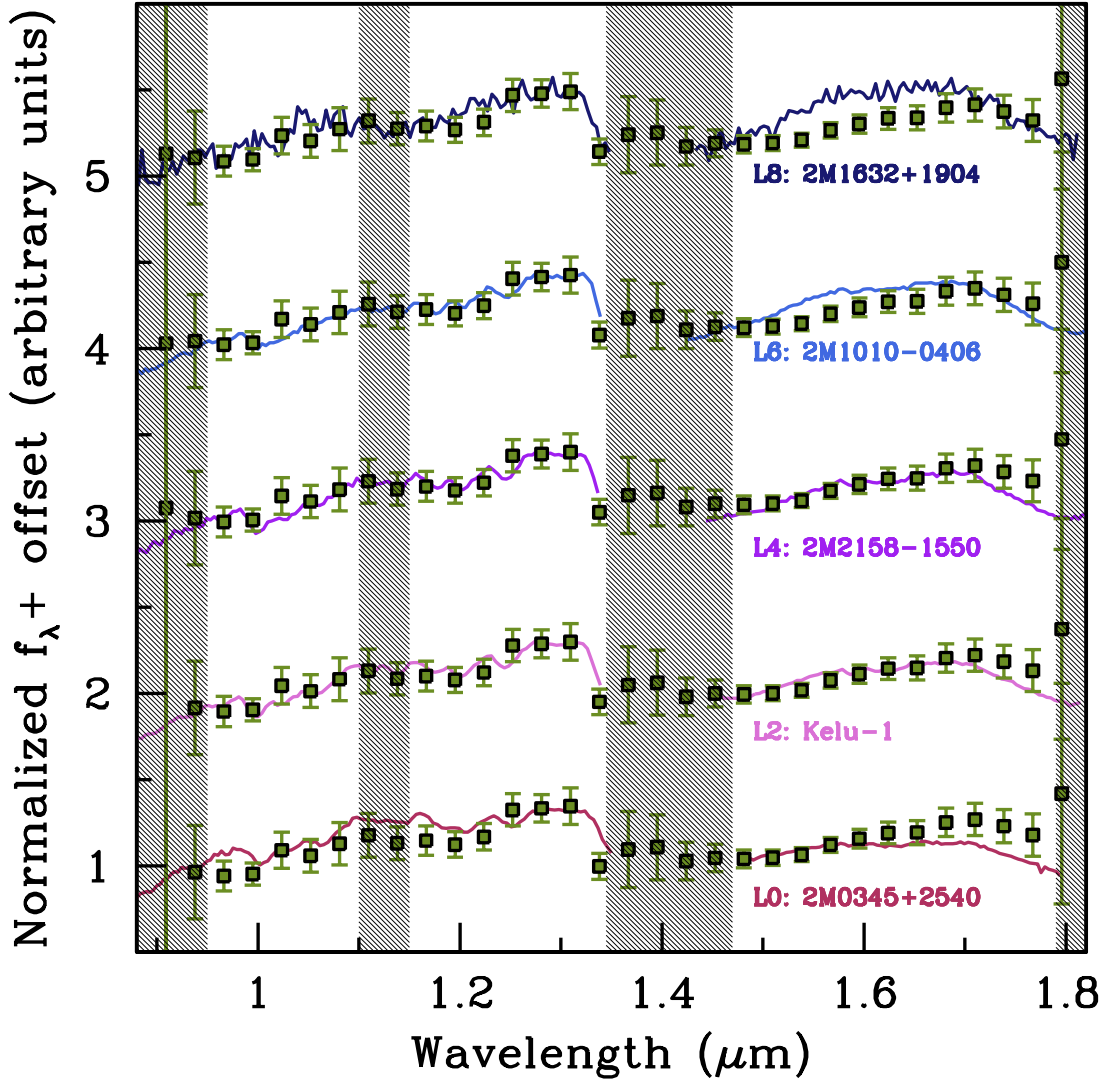}
\end{center}
\caption{Comparison of Project 1640 spectra of $\kappa$ And B (points)
  with field brown dwarf standards ranging from L0 to L8 taken from
  \citet{klb10}.  Each spectrum has been offset arbitrarily for
  clarity, although no other offsets of any kind have been applied to
  either the template spectra or the P1640 data. The shaded regions
  indicate spectral regions where telluric water absorption is strong.
}
\label{fig:fieldBDs}
\end{figure*}

In this section, we present new spectrophotometry of $\kappa$ And B
which is compared with empirical and synthetic spectra of substellar
objects, as well as an analysis of the near-infrared colors of the
object.

\begin{deluxetable}{llll}
\tabletypesize{\scriptsize}
\setlength{\tabcolsep}{0.03in}
\tablewidth{0pt}
\tablecaption{Derived Properties for $\kappa$ And B \label{tab:companion_props}}
\tablehead{
\colhead{Parameter} & Value & Units & \colhead{Reference} 
}
\startdata
\teff                         &        2040$\pm$60      &  K               &  This work (\S\ref{sec:synthetic})\\
Spectral Type      &       L1$\pm$1               &                    &  This work (\S\ref{sec:p1640_spectra})\\
Mass                     &        50$^{+16}_{-13}$& \mjup         &  This work (\S\ref{sec:synthetic})  \\
\logg                      &         4.33$^{+0.88}_{-0.79}$  &            &  This work (\S\ref{sec:synthetic})\\
Age                        &  220$\pm$100             &   Myr          &   This work (\S\ref{sec:age})
\enddata
\tablecomments{}
\end{deluxetable}

\subsection{Spectroscopy from 0.9 - 1.8 $\mu$m}\label{sec:p1640_spectra}

We imaged the $\kappa$ And system on UT 2012 December 23 using
``Project 1640'' \citep[][]{hoz11,obb12} on the 200-in Hale Telescope
at Palomar Observatory. Project 1640 is a coronagraph integrated with
an integral field spectrograph \citep[``IFS'',][]{hob08} covering the
$YJH$-bands. This instrument ensemble is mounted on the Palomar
``PALM-3000'' AO system \citep{dbp98, rdb12}, which in turn is mounted
at the Cassegrain focus of the Hale Telescope.  In addition, the
system uses an internal wave front calibration interferometer
\citep[e.g.][]{wgs04,zvs12} for reducing non-common path wave front
errors internal to the instrument ensemble, thereby boosting
performance at small angular separations.

Starting at an airmass 1.02, sixteen Project 1640 multi-spectral
images were obtained, each with exposure time of 183s.  The star was
placed behind the coronagraphic mask, the PALM-3000 AO control loops
were locked, and additional corrective wave front sensor offsets were
applied to the PALM-3000 AO system from the wave front calibration
interferometer, thus minimizing the halo of correlated speckle noise.
To alleviate the inherent uncertainty in the position of the occulted
star in coronagraphic images \citep[e.g.][]{dho06}, the position of
the star was determined by using a set of fiducial reference spots
created by a physical pupil plane grid in the Project 1640 coronagraph
\citep{so06,mlm06}.

The Project 1640 data reduction pipeline is described in
\citet{zbo11}. To convert the image data counts obtained by the
spectrograph to physically meaningful quantities, the counts in a
Project 1640 pupil plane image of the $\kappa$ And system were
measured with the star moved off the coronagraphic mask obtained
shortly after the science observations. In this configuration, the
entire field of view of the pupil plane is uniformly illuminated.
Comparing the counts measured in each channel in the data cube with
the actual flux value from an empirical spectrum for a B9 star from
the Pickles Stellar Spectral flux library \citep{p98} provides a
relation of data counts in the science camera to physical units of
flux density.


\begin{figure*}
\begin{center}
\epsscale{1.0}
\plotone{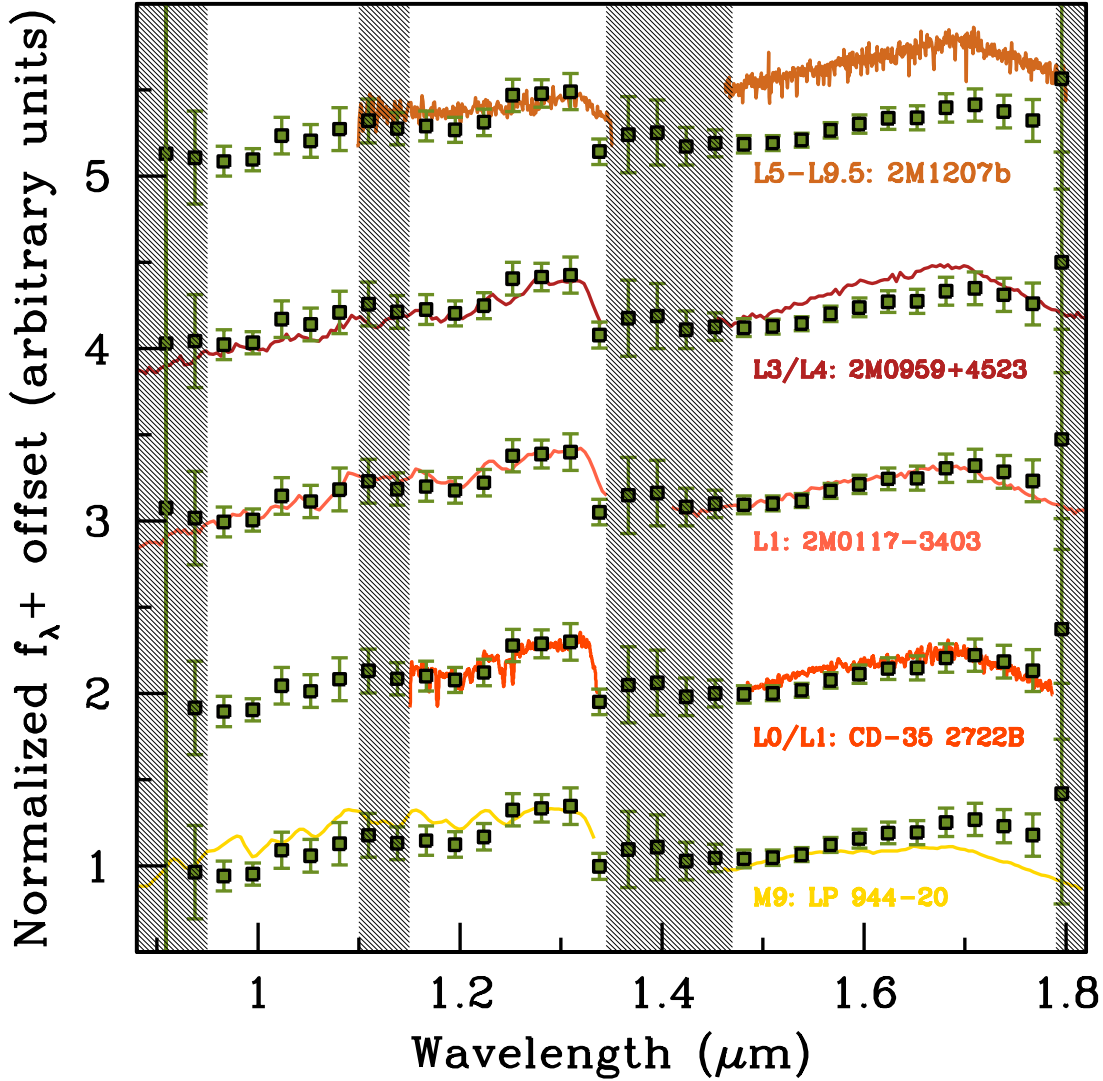}
\end{center}
\caption{ Comparison of the Project 1640 spectra of $\kappa$ And B
  (green points) with several young and intermediate age L-dwarfs
  ranging from L1 to later than L5 (see text for details).  Also shown
  is an M9 object presented in \citet{bli08} as well as the very young
  2M1207b \citep{pkd10}.  The best match empirical spectra to the
  Project 1640 data is the $\sim$50-150 Myr L1$\pm$1 object
  2M0117-3403 (Faherty et al. {\it in prep}), which has a
  $(J-H)$=0.97$\pm$0.05 color, consistent with the published value of
  0.91 $\pm$ 0.1 from \citet{bcm13}.  Each spectrum has been offset
  arbitrarily for clarity, although no other offsets of any kind have
  been applied to either the empirical spectra or the Palomar
  data. The shaded regions roughly indicate spectral regions where
  telluric water absorption is strong.  }
\label{fig:youngBDs}
\end{figure*}

Extracting a spectrum of an object such as $\kappa$ And B is
challenging due to the $\sim$10$^4$ contrast ratio between it and the
host star at only 1$^{\prime\prime}$.  The single largest hindrance to
extraction of high signal-to-noise spectra is the quasi-static speckle
noise in the image focal plane \citep{rwn99,mdr00,hos07}.  For objects
with brightness greater than, or comparable to the speckle noise halo
\citep[e.g.][]{hob10,zoh10,hmo11,phv12,hho13}, evaluation of companion
spectra can be performed with conventional aperture photometry.
However, for objects with higher contrast such as $\kappa$ And B or HR
8799 \citep{obb13}, an IFS can improve sensitivity through the
suppression of this quasi-static speckle noise in the image
\citep{cpb11,pcv12}.



To reduce the effects of quasi-static speckle noise in our
multi-spectral images, we use speckle suppression techniques based on
the principle component analysis algorithm outlined in
\citet{spl12}. This method uses a basis of eigenimages created by a
Karhunen-Lo\`eve transform of Point Spread Function (PSF) Reference
images to perform the PSF subtraction, and is amenable to point source
forward modelling (Pueyo et al {\it in prep.}). Applications of this
method have been demonstrated for the directly imaged planets in the
HR 8799 system \citep{obb13}.
Figure~\ref{fig:P1640_s4} shows three images from the Project 1640 IFS
subsequent to our speckle suppression post-processing corresponding
the $Y$, $J$, and $H$-band central wavelengths.

As with any form of PSF subtraction \citep[e.g. classical ADI,
LOCI,][]{mld06,lmd07}, the performance of the algorithm centers on
robust co-alignment of the PSF reference images.  To align our
reference images, we perform an initial alignment based on the
fiducial astrometric reference spots in the data, and perform a
subsequent sub-pixel cross-correlation using the image speckles.  To
verify that the extraction of spectra is robust and that flux is not
significantly depleted from the $\kappa$ And B source, we execute a
parameter space search as follows: we use 16 different geometries near
the location of the companion (these include varying the size of the
search zone and the radial exclusion parameter necessary to mitigate
cross talk between nearby spectral channels), and vary the number of
eigenmodes in the PSF subtraction from 1 to 130.  This parameter
search results in $\sim$2000 spectra for the $\kappa$ And B companion.
From these $\sim$2000 spectra, we discard any spectra for which a) the
astrometric position of the companion is not consistent between
wavelength channels; b) a sharp flux drop is detected with a small
change in the number of modes.
Eliminating reduced spectra in case a) ensures that the extraction is
not biased by residual speckle noise (should a speckle have some
overlap with the companion then it will yield a wavelength dependent
astrometric bias). A sharp drop in flux over a small number of modes
indicates that too many eigenmodes are being used, thus in case b) we
eliminate spectra corresponding to overly aggressive PSF
subtractions. Finally we further trim this subset by only keeping the
spectra that exhibit a {\em local} SNR $>$ 3 for all wavelengths in
the band of interest (either $Y+J$ or $H$). This procedure leads to
$\sim$40 high quality spectra for the companion. The mean of these
values comprises the spectral points plotted in
Figures~\ref{fig:fieldBDs} and \ref{fig:youngBDs} (green individual
points), and the error bars denote the standard deviation of these
$\sim$40 spectra, plus the uncertainty associated with the pupil-plane
spectral calibration added in quadrature.



\begin{figure}
\begin{center}
\epsscale{1.2}
\plotone{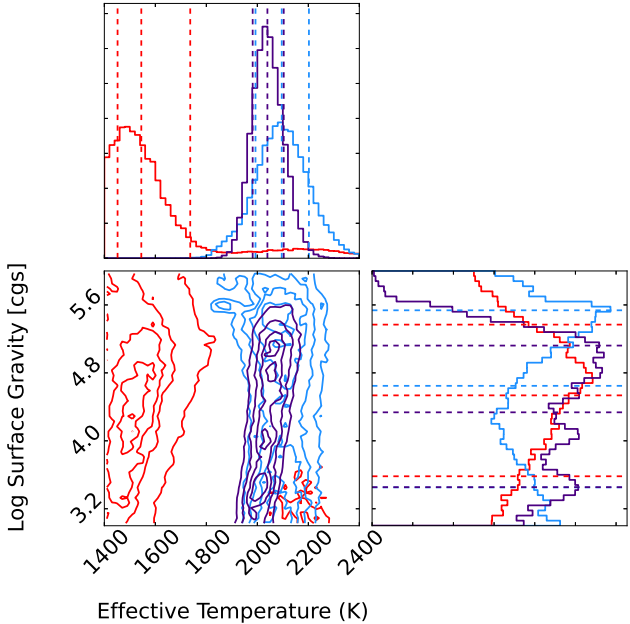}
\end{center}
\caption{Posterior distributions for the MCMC fits of synthetic
  spectra to the Project 1640 data shown in Figure~\ref{synthetic}.
  The figure uses the same color scheme as
  Figure~\ref{synthetic}. Namely, the blue distributions show the
  results to the fits of only the $Y$ and $J$ portions of the spectra,
  red corresponds to $H$-band, and purple is the full spectral range
  $YJH$. The best-estimate for each parameter correspond to the 50\% quantile.  One dimensional representations of the marginalized temperature and
  surface gravity posterior distributions are shown at the top and right,
  respectively, with best-estimate, and 68\% confidence intervals marked by the dashed lines.}
\label{fig:mcmc}
\end{figure}

\subsection{Comparison with Empirical Brown Dwarf Spectra}\label{sec:BDcomparison}

Figure~\ref{fig:fieldBDs} shows the $YJH$-band spectro-photometry from
Project 1640 for $\kappa$ And B.  Overlaid with the spectra in
Figure~\ref{fig:fieldBDs} are empirical spectra for field-age
``standard'' objects taken from \citet{klb10}, ranging from spectral
type L0 to L8. The empirical spectra are normalized such that they
match the Project 1640 flux at 1.28 $\mu$m, and the figure identifies
three regions near 1.1, 1.4, and 1.8 $\mu$m where telluric water
absorption is particularly strong.
Assuming a surface gravity value appropriate for dwarf-like objects,
the Palomar data show a best match to the mid-L spectral types.
Specifically, as demonstrated in Table~\ref{tab:chi_values}, a
$\chi^2$ goodness-of-fit metric reveals a best-fit to the L4 field
object 2MASS~J21580457-1550098.


However, as we discuss in \S\ref{sec:age}, the $\kappa$ And system
likely has an age of 220$\pm$100 Myr, significantly younger than
typical field ages ($\sim$few Gyr).  Thus, comparison with spectra of
L-dwarfs with known indicators of youth may be more appropriate.
Figure~\ref{fig:youngBDs} shows our $YJH$-band spectro-photometry
along with several young and intermediate age substellar objects
ranging from the very young ($\sim$10 Myr) object 2MASS
J12073346-3932539b
(hereafter, 2M1207b)
\citep[][]{pkd10} to the mid-late young L-dwarf
2M0959 ($\sim$150 Myr, Faherty et al. {\it in prep}), and LP~944-20, a
$\sim$300 Myr late-M dwarf \citep{bli08}. Several of these objects are
discussed in \S\ref{sec:nir_photometry} and shown in
Figure~\ref{fig:jmk}, as well.  Among the young objects, the best
fitting ($\chi^2$=0.76) synthetic spectrum to our data is
2MASS~J01174748-3403258 (hereafter, 2M0117-3403), a L1$\pm$1
``$\beta$'' intermediate-gravity brown dwarf \citep{al13}.  Indeed,
the best fit template spectrum of the L1$\pm$ object 2M0117 to our
$\kappa$ And B spectra has a $(J-H)$= 0.97$\pm$0.05 well matched to
that of $\kappa$ And B. Thus, comparing to objects with ages more
appropriate to $\kappa$ And implies a slightly earlier spectral type
object \citep[L1: 2000$\pm$200K, e.g. ][]{k05} than would be inferred
for a later type field age L4 object (1700-1900K).

\begin{figure}[ht]
\center
\resizebox{1.0\hsize}{!}{\includegraphics[angle=0]{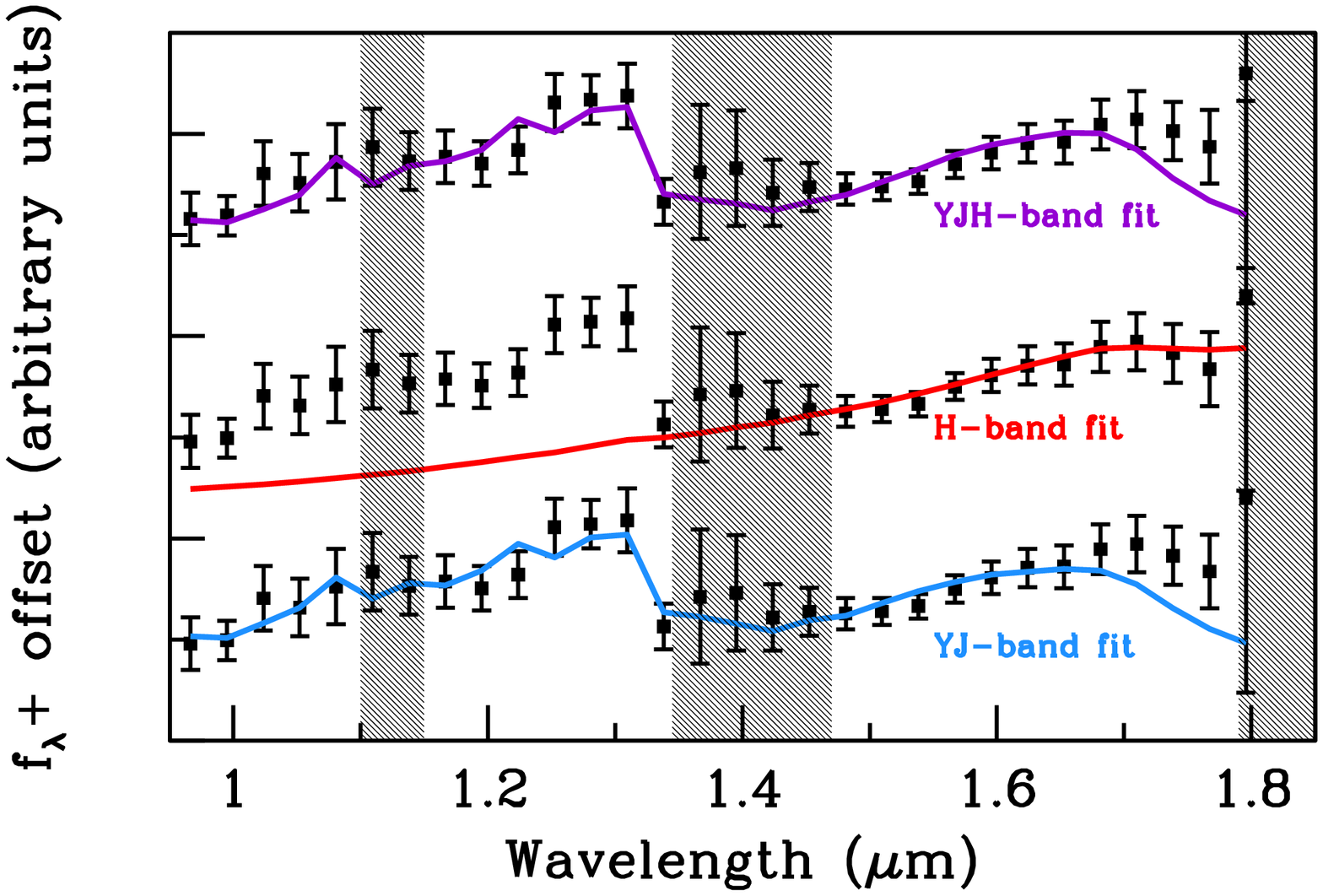}}
 \caption{The best fit synthetic PHOENIX models \citep{hba97,bha01, aha01} to the Project 1640 spectra. The models have been fit using the methods of \citet[][]{rrb12}  and Rice et al. ({\it in prep}). The lower curve (blue) is the best-fit synthetic model to
   only the $Y$ and $J$-band Project 1640 points, while the middle
   curve (red) shows the best fit synthetic model for $\kappa$ And B
   to only the $H$-band spectral points (1.45-1.80 $\mu$m). The top
   curve (purple) reflects the best fit values for all the $YJH$-bands
   simultaneously. 
   Figures~\ref{fig:mcmc} and \ref{Rice_figure} use the same color scheme. 
   }
  \label{synthetic} 
\end{figure}

The comparable quality of fits to empirical spectra of the
intermediate-gravity L1 object as well as the L4 field object prevents
us from placing extremely strong constraints on the spectral type of
$\kappa$ And B.  Given the relatively low spectral resolution and
finite wavelength coverage ($YJH$-bands) Project 1640 data, these data
may only be able to discern a range of spectral types for this
companion (e.g.~$\sim$L1-L4). Further, discerning gravity-sensitive
features from these data may be challenging, especially at high
contrast, and when heavy contamination from speckle noise is present.
Observations of targets with well known gravity features may be needed
to calibrate the strength of gravity effects in the data. Nonetheless,
our best match to the low-gravity young object 2M0117 is likely a
point of consistency with our revised 220 Myr age of the primary
(\S\ref{sec:age}).
We thus
adopt a conservative estimate of L1$\pm$1 for $\kappa$ And B.  As we
show below, fitting synthetic models to our spectra give temperatures
consistent with an L1$\pm$1 object (\S\ref{sec:synthetic}), and
color-magnitude diagram analysis supports the L1$\pm$1 identification
(\S\ref{sec:nir_photometry}, Figure~\ref{fig:colormag}).

\begin{deluxetable}{llll}
\tabletypesize{\scriptsize}
\setlength{\tabcolsep}{0.03in}
\tablewidth{0pt}
\tablecaption{$\chi^2$ Goodness of Fit values for Brown Dwarf Template Spectra \label{tab:chi_values}}
\tablehead{
\colhead{} & \colhead{Object} & \colhead{Spectral Type} &  \colhead{$\chi^2$} 
}
\startdata
Young Brown Dwarfs: &           &                    & \\
&  LP~944-20           &    M9          & 1.91        \\
& CD-35~2722 B      &    L0-L1      &  4.03 \\ 
& 2M~0117-3403     &    L1           & 0.76 \\
& 2M~0959+4523     &    L3-L4      & 1.97 \\
& 2M~1207b             &     L5-L9.5  &  $>$40 \\ \hline 
Field objects: &          &                     &            \\
& 2M~0345+2540           &     L0            &  1.81 \\
& Kelu-1                          &     L2            &  0.96 \\ 
& 2M~2158-1550             &    L4             & 0.65 \\ 
& 2M~1010-0406             &     L6             & 1.91 \\
& 2M~1632+1904            &    L8              & 4.15 
\enddata
\tablecomments{}
\end{deluxetable}




\subsection{Comparison with Synthetic Spectra}\label{sec:synthetic}

To directly constrain the physical properties of this object, we
compare \citep[e.g.][, Rice et al. {\it in prep}]{rrb12} the observed P1640 spectrum to a grid of synthetic spectra
from the PHOENIX models \citep{hba97,bha01, aha01}. The model
atmospheres cover \teff\ = 1400K to 4500K and \logg = 3.0-6.0 in
intervals of 50K and 0.1 dex at solar metallicity using the {\it
  dusty} version of dust treatment and are described in more detail by
\citet{rbm10} and \citet{rrb12}. The adopted best-fit parameters are the 50\% quantile values of the $10^6$ link posterior distribution functions from a Markov chain Monte Carlo
(MCMC) analysis that interpolates between calculated synthetic
spectra, creating an effectively continuous grid of models.

\begin{figure}
\center
\epsscale{1.20}
\plotone{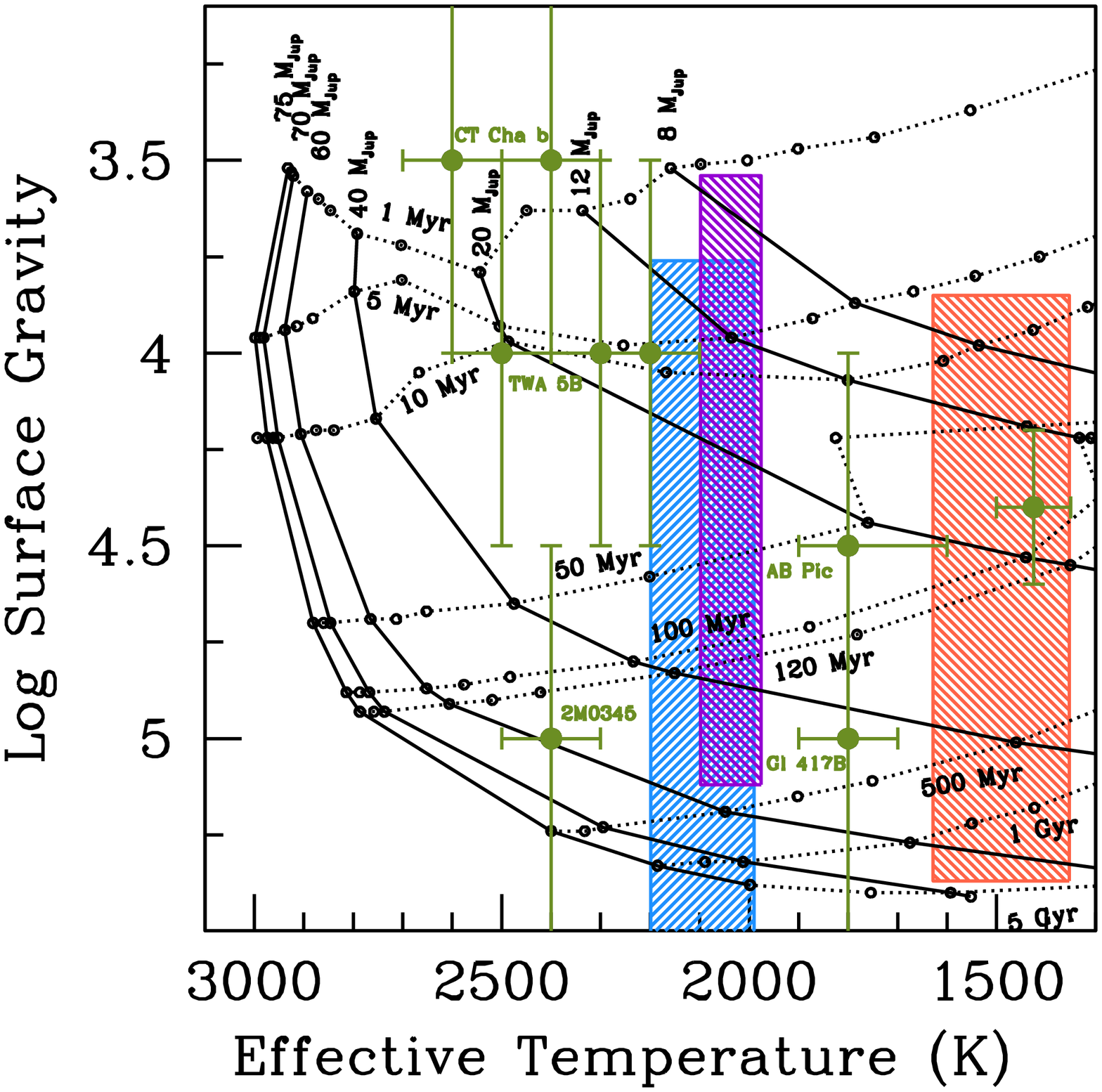}
\caption{The best fit \logg~and \teff\ values for $\kappa$ And B
  derived by comparing to sythetic spectra using the methods of
  \citet{rbm10}. The red box identifies the $\pm$68\% confidence intervals 
  of the fit in both temperature and gravity to only the Project 1640 H-band spectral points
  (1.45-1.80 $\mu$m) for $\kappa$ And B, the blue region indicates the $\pm$68\% confidence intervals for the
  $YJ$-bands, while the purple region reflects the $\pm$68\% confidence intervals
  for all the Project 1640 wavelengths simultaneously ($YJH$-bands).
  The best fit synthetic specta are shown in Figure~\ref{synthetic},
  and the posterior distributions for our MCMC fitting procedure are
  shown in Figure~\ref{fig:mcmc}.  Also shown are age/mass isochrones
  from the DUSTY00 models from \citet{cba00} and \citet{bca02}. The
  gray circular points indicate the \logg~and \teff\ values from
  several young, low mass M and L-type objects taken from
  \citet{bcl13} and \citet{bls13}.
  }
  \label{Rice_figure} 
\end{figure}

We use three different spectral ranges: 0.9 - 1.32$\mu$m ($YJ$-bands),
1.47-1.78$\mu$m ($H$-band), and the full range 0.9 - 1.78$\mu$m
($YJH$-bands) for the spectral comparison. We exclude the four points occupying the water band separating the $J$ and $H$-bands ($\sim$1.4$\mu$m) as well as the final $H$-band point near 1.8$\mu$m, but we include the two points in the water band between the $Y$ and $J$-bands, since their uncertainties are comparable with points in the bandpasses.   Figure~\ref{synthetic}
shows the three best fit synthetic spectra and their input physical
parameters for each spectral region, plotted over the entire observed
spectrum. The fit to the full set of data ($YJH$-bands) is the best
overall and has similar parameters (2040K$^{+58}_{-64}$, \logg=4.33$^{+0.88}_{-0.79}$, 
$\pm$68\% confidence intervals )
as the $YJ$-band only fit (2096K$^{+103}_{-106}$, \logg=4.65$^{+1.20}_{-0.89}$).  As
Figure~\ref{synthetic} shows, the fit to only the $H$-band data
significantly underpredicts several flux points in $YJ$-bands,
producing a temperature $\sim$1550K, and we regard this as physically
unreasonable.  Figure~\ref{fig:mcmc} shows the posterior distributions
for fits for each of the three cases, along with the margnalized one-dimensional posterior distributions for each fitting parameter.  


\begin{figure}
\begin{center}
\epsscale{1.15}
\plotone{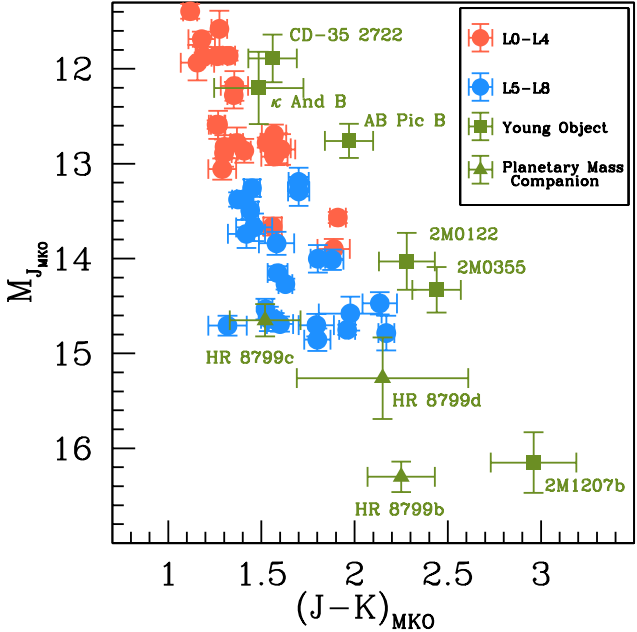}
\end{center}
\caption{The near-infrared color magnitude diagram for well-studied
  field and young L-type brown dwarfs as well as planetary mass
  companions. Absolute magnitudes were derived from parallaxes
  reported in \citet{Faherty12} and \citet{Dupuy12}.  The placement of
  $\kappa$ And B is consistent with an L1$\pm$1 spectral type.  We
  also show comparably young L dwarfs 2M0355, AB Pic b, and CD-35~2722B, and 2M0122-2439B (See also Figure~\ref{fig:youngBDs} and
  Figure~\ref{fig:jmk}).}
\label{fig:colormag}
\end{figure}

In Figure~\ref{Rice_figure} we compare the best-fit \teff~and
\logg~values to parameters predicted by DUSTY00 models from
\citet{cba00} and \citet{bca02} for ages ranging from 1 Myr to 5 Gyr,
and masses ranging from 8 \mjup\ to 75 \mjup. The uncertainties on the
best-fit physical parameters for $\kappa$ And B represent the width of
the distribution of the 10$^6$-link Markov chain values marginalized
over the other parameter, as described in \citet[][Rice et al. {\it in
  prep.}]{rrb12}. Also shown are the locations in \logg~versus
\teff~space of several young M and L-type objects taken from
\citet{bcl13} and \citet{bls13}.  

The locations of the $YJ$-band and $YJH$-band best fit \teff\ values
(2000-2100K) and their uncertainties are consistent with an L1$\pm$1
spectral type \citep[e.g.][]{k05, slc09}. 
However, the constraints on the surface gravity for $\kappa$ And B still permit a wide range of ages, from very young to several hundred Myr.  However, the range still includes our revised age of 220$\pm$100 Myr. 
Indeed, low resolution spectral fits to known young very low
mass objects presented in Rice et al. ({\it in prep.})  also suggest
surface gravities higher than would be expected for the
$\sim$10--100~Myr ages, possibly indicating the inadequacy of the
simplified dust treatment of the {\it dusty} PHOENIX model atmospheres
in recreating the emergent spectra of young, very low mass objects.

\begin{figure}
\begin{center}
\epsscale{1.15}
\plotone{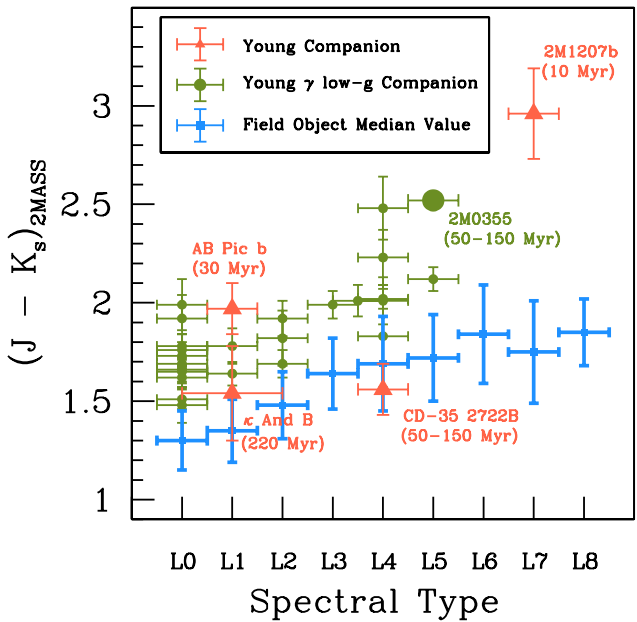}
\end{center}
\caption{2MASS ($J-K_{s}$) color versus spectral type for field L0-L9
  dwarfs. Mean colors of normal (excluding subdwarfs and suspected
  young) objects are displayed as blue points with error bars.  Low
  surface gravity $\gamma$ L-dwarfs (denoted by ``$\gamma$ Low-G'')
  are the grey points with the most extreme, 2M0355, highlighted.
  Companion L dwarfs are shown as red triangles. Using the ($J-K_{s}$)
  color as a coarse age discriminator, $\kappa$ And B is evidently
  older than the very young objects such as AB Pic b, 2M1207b, and
  2M0355, but still possibly consistent with the population of
  $\gamma$ L-dwarf.s}
\label{fig:jmk}
\end{figure}

\subsection{Near-infrared Luminosity and Colors}\label{sec:nir_photometry}

Combining luminosity with NIR color has emerged as a potentially
powerful lever for deciphering age properties of brown dwarfs and
giant planets.  For example, normal field L dwarfs typically have
$(J-K_s)$=1.3-1.8, while the $\gamma$ low-gravity sources are
$\sim$0.3-0.6 magnitudes redder than the median for their respective
spectral types (Figure~\ref{fig:jmk}). When one combines the absolute
$JHK$ magnitudes of the $\gamma$ sources and compares them to
equivalent spectral type targets, one finds they are not just redward
but also up to 1.0 mag underluminous \citep{Faherty12, Faherty13} in
the NIR.  The same trend has been cited in directly imaged giant
exoplanet studies (e.g. 2M1207b and HR8799b; Chauvin et al. 2004;
Marois et al. 2010).  Figures~\ref{fig:colormag} and ~\ref{fig:jmk}
show the properties of $\kappa$ And B compared to field brown dwarfs
as well as four well studied, young L dwarfs: AB Pic B, CD-35~2722 B,
2M0355+1133, and 2M1207 B \citep{clz05,wlb11,Faherty13,cld04}.
Several of these young L-dwarfs are used for comparison in
Figure~\ref{fig:youngBDs} (\S\ref{sec:p1640_spectra}).

As shown in Figure~\ref{fig:colormag}, $\kappa$ And B has a comparable
absolute magnitude to L0-L4 dwarfs, including the planetary mass
object AB Pic B (SpT L1; Bonnefoy et al. 2013).  However, AB Pic B is
redder than the ``main sequence" of L and T dwarfs as well as $\kappa$
And B, and forms a sequence with comparably young sources 2M1207b, and
2M0355. Indeed, Figure~\ref{fig:jmk} shows that $\kappa$ And B is
consistent with the median $(J-K_s)$ colors of field L-dwarfs, and is
marginally inconsistent with the young $\gamma$ low-gravity objects.
The luminosity alone rules out $\sim$mid to late spectral types as the
lower temperatures would make $\kappa$ And B significantly
overluminous. This suggests an earlier spectral type, consistent with
the L1$\pm$1 spectral type derived in \S\ref{sec:BDcomparison}.

\section{Properties of $\kappa$ And A}\label{sec:props}

In \S\ref{sec:secondary_props} we presented spectroscopic and
photometric evidence that the companion to $\kappa$ And is more 
consistent with an object that is older and higher mass than the young (30 Myr), and low mass (12-14
\mjup) that has been claimed in the literature.  In this section, we
present further constraints on the age of the system through an
analysis of fundamental properties of the host star.

\subsection{Stellar Parameters}\label{sec:stellar_params}

$\kappa$ And A is a V = 4.138\,$\pm$\,0.003 mag \citep{Mermilliod91}
B9IVn \citep{Cowley69, GrayGarrison94} star at distance
51.6\,$\pm$\,0.5 pc \citep[$\varpi$ = 19.37\,$\pm$\,0.19
mas;][]{vanLeeuwen07}.  The ``n'' in the spectral type signifies that
it is a fast rotator, however its \vsini~\citep[150 \kms\,; ][]{Abt02}
is not unusual for field B9 stars \citep{Kraft67}. At this distance,
results from reddening surveys suggest that the star should be
negligibly reddened \citep[$E(b-y) <$ 0.02 mag ;][]{Reis11}.


Several lines of evidence point to a lower surface gravity for
$\kappa$ And A compared to what would be expected of a $\sim$30
Myr-old star. Early indications of a luminosity class different than
the dwarf categorization were presented by \citet{Cowley69}, and later
by \citet{cmj77}, whose ultraviolet line analysis led to
classification of $\kappa$ And A as ``gB9'', indicating a surface
gravity more indicative of giants rather than dwarf stars.
Thereafter, fitting of atmospheric models to ultraviolet photometry by
\citet{Malagnini83} derived a surface gravity of \logg\,=\,3.69 --
lower than the \logg$\simeq$4.2-4.5 value typical for luminosity class
V stars.
Further, surface gravity estimates of \logg\, = 4.17, 4.10, 3.97,
3.87, and 3.78 were estimated by \citet{AllendePrieto99},
\citet{Fitzpatrick05}, \citet{Prugniel07}, \citet{Wu11}, and
\citet{bcm13} respectively. Taken as a group, these values are more
commensurate with luminosity class IV than typical class V dwarf
stars.

\begin{figure*}
 \epsscale{1.17}
\plottwo{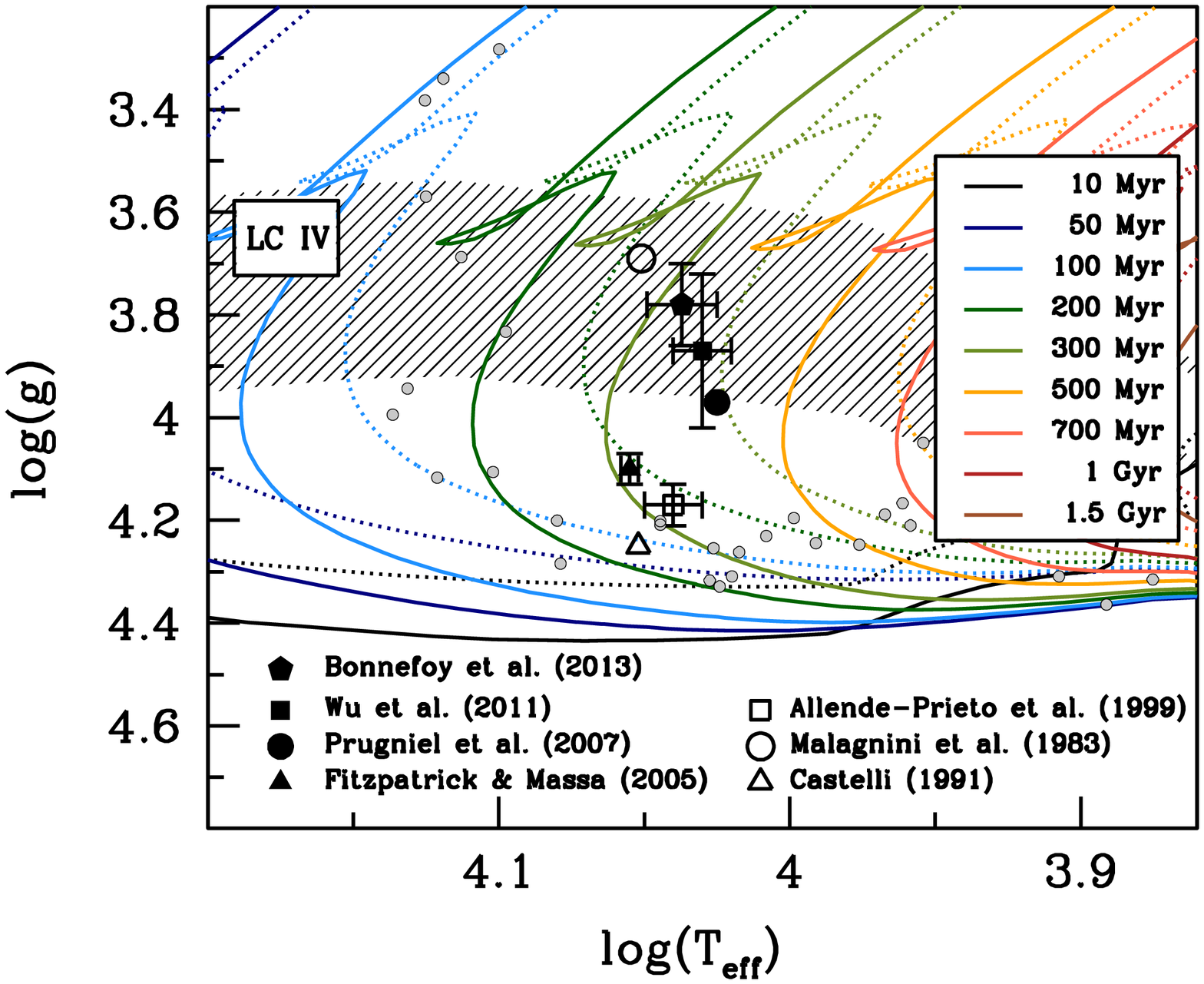}{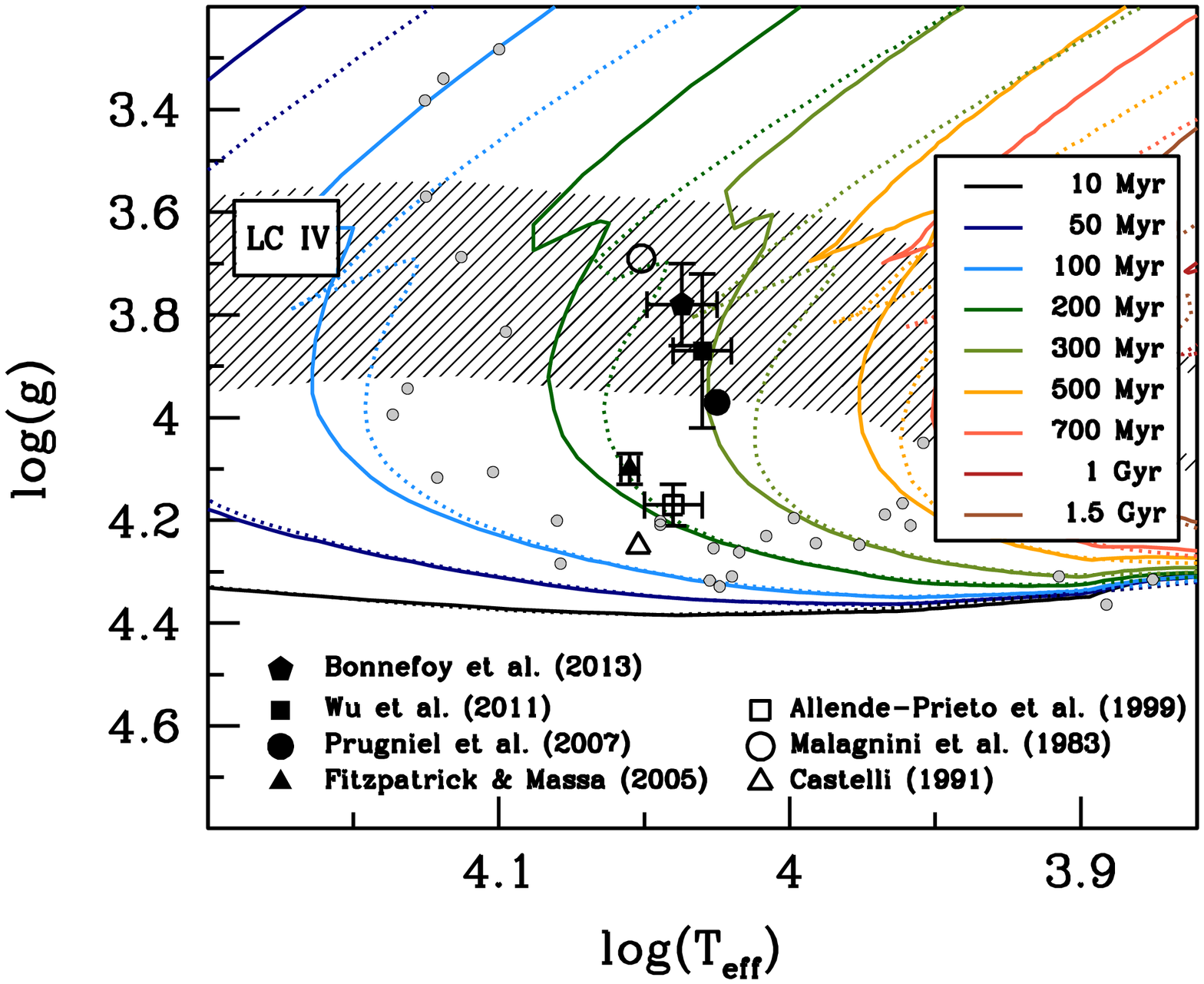}
\caption{ Previously published (see \S\ref{sec:age}) determinations of
  \teff\ and \logg\ for $\kappa$ And (large points) are compared with
  Pleiades members as discussed by David et al. (2014, {\it in
    prep}). {\it Left:} overlaid are the PARSEC isochrones of
  \citet{bmg12}. The solid isochrones are for a metallicity of
  [M/H]=-0.36, the value for $\kappa$ And determined by
  \citet{Fitzpatrick05}.  The dashed isochrones are for solar
  metallicity. The isochrone ages include the pre-main sequence
  evolution timescales. All of the published determinations of
  \teff~and \logg~for $\kappa$ And are consistent with an isochrone
  age $>200$ Myr in the sub-solar metallicity case and an age $>50$
  Myr in the solar metallicity case.  The shaded band labeled ``LC
  IV'' identifies the range of spectroscopic \logg\ measurements
  occupied by subgiant standard stars taken from the PASTEL data base
  (see text).  {\it Right:} The solid curves are isochrones of
  \citet{ege12} computed from stellar evolutionary models that start
  on the ZAMS with a rotation rate of $v_{rot}/v_{crit} = 0.4$. The
  dashed curves are isochrones computed from stellar evolutionary
  models with zero rotation. All of the published determinations of
  \teff\ and \logg~are consistent with an isochrone age $>100$ Myr for
  $\kappa$ And, with several being consistent with $\sim$200 Myr. }
 \label{fig:trevor_figure}
 \end{figure*}

As a demonstration of this, Figure~\ref{fig:trevor_figure} shows these
values from the literature in a \logg\ versus \teff\ diagram. This
figure also shows a region of \logg\ space (shaded band) spanned by a
compendium of spectroscopic surface gravity measurements for subgiant
standard stars from the PASTEL database
\citep{Soubiran10}\footnote{http://vizier.u-strasbg.fr/viz-bin/VizieR?-source=B/pastel}.
An assessment of the quality of the spectroscopic standards was made:
those that showed the most consistency in the literature, and/or had
the best pedigrees were used.  This band has a width of $\pm$0.19 dex,
as determined by the 1$\sigma$ variation of the \logg\ measurements
from the subgiant standards.  This scatter reflects a mix of
differences amongst spectroscopic \logg\ values published for a single
subgiant standard star, and standard-to-standard differences. Given
the heterogeneity of the data, we do not attempt to disentangle which
effects dominate, but the data seem to suggest that $\pm$0.2 dex rms
accuracy in \logg\ is a reasonable lower limit on the predictive power
of luminosity class IV to predict surface gravity.  Fitting a
fifth-order polynomial to the \logg measurements (spanning spectral
types B0 to K1) predicts a \logg=3.75$\pm$0.19 for a B9IV star.


 $\kappa$ And's effective temperature (\teff) has been estimated as
 10594\,K \citep{Prugniel07}, 10733\,$\pm$\,247\,K \citep{Wu11},
 10839\,$\pm$\,200\,K \citep{Zorec12}, 10965\,K
 \citep{AllendePrieto99}, 11246\, K \citep{g78}, 11240\,K
 \citep{Malagnini83, Morossi85}, 11310\,K \citep{Napiwotski93},
 11361\,$\pm$\,66 K \citep{Fitzpatrick05}, 11535\,K \citep{Westin85}.
 Arguably the most comprehensive analysis is by \citet{Fitzpatrick05},
 who fit Kurucz ATLAS9 model synthetic spectra to optical/infrared
 photometry and {\it IUE} ultraviolet spectra. \citet{Fitzpatrick05}
 derived extremely well-constrained stellar parameters of \teff\, =
 11361\,$\pm$\,66\,K, [Fe/H] = -0.36\,$\pm$\,0.09 dex, radius
 2.31\,$\pm$\,0.09 R$_{\odot}$, and spectroscopic surface gravity
 log($g$) = 4.10\,$\pm$\,0.03 dex. Given their adopted distance based
 on the original Hipparcos catalog ($d$ = 52.0\,$\pm$\,1.6 pc), their
 parameters imply an angular diameter of 413\,$\pm$\,16 $\mu$as (of
 which $\pm$13 $\mu$as is due to the distance error, with presumably
 $\pm$10 $\mu$as coming from the uncertainty in the bolometric flux).
 Using the revised Hipparcos parallax of $\varpi$ = 19.37\,$\pm$\,0.19
 mas ($d$ = 51.63\,$\pm$\,0.51 pc; 1\%\, error), we update the
 luminosity estimate from \citet{Fitzpatrick05} to log($L/L_{\odot}$)
 = 1.895\,$\pm$\,0.024 dex.

 More accurate determinations of luminosity and \teff\ could easily be made if the inclination of $\kappa$ And were to be determined through interferometry \citep[e.g.][]{Monnier12}, enabling stellar parameters to be computed that are not biased by our unknown viewing angle of  this rapid rotator.  
 We discuss aspects of the star's inclination in greater detail in \S~\ref{sec:inclination}. 



\begin{deluxetable}{lccccc}
\tabletypesize{\scriptsize}
\setlength{\tabcolsep}{0.03in}
\tablewidth{0pt}
\tablecaption{Stellar Parameters\label{tab:data}}
\tablehead{
{(1)}           &{(2)}   &{(3)}   & {(4)}\\
{Parameter}     &{Value} &{Units} & {Ref.}
}
\startdata
Parallax($\varpi$)   &  19.37\,$\pm$\,0.19 & mas    & 1\\
Distance(1/$\varpi$) &  51.63\,$\pm$\,0.51 & pc     & 1\\
\pmra\,          &  80.73\,$\pm$\,0.14 & \masyr & 1\\
\pmdec\,         & -18.70\,$\pm$\,0.15 & \masyr & 1\\
\vrad\,          & -12.7\,$\pm$\,0.8   & \kms   & 2\\
U               & -11.5\,$\pm$\,0.3    & \kms   & this work\\
V               & -20.1\,$\pm$\,0.5    & \kms   & this work\\
W               & -5.9\,$\pm$\,0.6     & \kms   & this work\\
m$_V$            &  4.138\,$\pm$\,0.003 & mag   & 3\\
M$_V$            &  0.574\,$\pm$\,0.037 & mag   & 4\\
\teff,          &  11361\,$\pm$\,66   & K     & 5\\
\vsini\,         & 150                  & \kms & 6\\
\logl\,         & 1.895\,$\pm$\,0.024 & dex & 7\\
Radius          & 2.29\,$\pm$\,0.06  & \rsun & 7\\
Mass            & 2.8$^{+0.1}_{-0.2}$  & \msun & this work\\
Age             & 220\,$\pm$\,100    & Myr & this work\\
\enddata
\tablecomments{References: 
(1) \citet{vanLeeuwen07},
(2) \citet{Gontcharov06},
(3) \citet{Mermilliod91}, 
(4) this paper, calculated using data in this table,
(5) \citet{Fitzpatrick05},
(6) \citet{Abt02},
(7) calculated using values from \citet{Fitzpatrick05}, updated
using the new Hipparcos parallax from \citet{vanLeeuwen07}.
}
\end{deluxetable}

\subsection{Age}\label{sec:age}

\subsubsection{Chemical Composition}\label{sec:chem}

The chemical composition of $\kappa$ And is worth briefly discussing
before trying to constrain its age using modern stellar evolutionary
tracks. Subsolar photospheric metallicities of [Fe/H] = -0.40
\citep{Prugniel07}, -0.45 \citep{Katz11}, -0.36 \citep{Fitzpatrick05},
and -0.32\,$\pm$\,0.15 \citep{Wu11} have been reported for $\kappa$
And. However, nearby, young ($<$200 Myr) open clusters and in the
solar vicinity have [Fe/H] $\sim$ 0.0 with rms scatter $\sim$0.1 dex
\citep[e.g.][]{Chen03}, as do the nearest star-forming regions ($<$10
Myr) and young stellar associations \citep{Santos08,VianaAlmeida09}.
If $\kappa$ And has kinematics consistent with a local origin, it is
highly unlikely that the bulk composition of $\kappa$ And would vary
significantly from solar.

For the Columba association, to which $\kappa$ And purportedly
belongs, the only stars from published membership lists
\citep{Torres08, Malo13, Zuckerman11, Zuckerman12} with spectroscopic
metallicity ([Fe/H]) estimates published in the PASTEL compendium of
stellar atmospheric parameters \citep{Soubiran10} are HD 984
\citep[0.09;][]{Valenti05}, HD 31647 \citep[-0.12;][]{Hill95}, HD
39206 \citep[0.06;][]{Lemke89}, and HD 40216
\citep[0.00;][]{Tagliaferri94}.  Hence, thus far, Columba members have
spectroscopic metallicities consistent with being approximately solar
(mean [Fe/H] = 0.01\,$\pm$\,0.05), and the spectroscopic [Fe/H]
estimates for $\kappa$ And would appear to make the star chemically
peculiar if it is truly associated with Columba.

Even {\it if} the bulk composition of $\kappa$ And is as metal poor as
the spectroscopic estimates listed above ([Fe/H] = -0.32 to -0.45)
indicate, this would only conspire to make the star systematically
older when comparing to evolutionary tracks.  In what follows, we
assume that $\kappa$ And has solar bulk composition, similar to other
very young stars in the solar vicinity. However, we also evaluate the
age assuming a lower metallicity.

\subsubsection{\logg\ versus \teff\ Analysis}\label{sec:logg_versus_teff}

%
As originally presented at IAU Symposium 299 in Victoria, BC on June
3, 2013, Figure~\ref{fig:trevor_figure} shows the \logg\ and \teff\
values previously listed in the literature plotted along with two sets
of isochrones for \logg\ and \teff.  The left plot shows the PARSEC
isochrones of \citet{bmg12} for two cases: a metallicity of
[M/H]=-0.36, the value for $\kappa$ \And\, determined by
\citet{Fitzpatrick05}, as well as solar metallicity. The isochrone
ages include the pre-main sequence evolution timescales. All of the
published determinations of \teff\ and \logg\ for $\kappa$ And are
consistent with an isochrone age $>200$ Myr in the sub-solar
metallicity case and an age $>50$ Myr in the solar metallicity case.

The right panel of Figure~\ref{fig:trevor_figure} shows the isochrones
taken from \citet{ege12}. These models are particularly applicable as
they take the effects of stellar rotation into account.  Indeed,
\citet{Carson13} use the work of \citet{ege12} to derive a stellar
mass.  Figure~\ref{fig:trevor_figure} shows the isochrones for a
rotation rate of $v_{rot}/v_{crit} = 0.4$
(See~\ref{sec:lum_versus_teff}), as well as those for zero
rotation. All of the published determinations of \teff\ and \logg~are
consistent with an isochrone age $>100$ Myr for $\kappa$ And, and
several values are consistent with the 200 Myr isochrone. Further, for
both plots, these literature points are located in a region of the
\logg\ versus \teff\ diagram where the isochrones are unambiguously
well separated.

Also shown in Figure~\ref{fig:trevor_figure} are several \logg and
\teff\ values taken for individual members of the Pleiades from the
$uvby\beta$ analysis of David et al. (2014, {\it in prep.}).  Each of
these points have had an individual $v\sin i$ rotation correction
factor applied to them to account for the rotation and inclination
effects discussed above.  These points show good agreement with the
solar metallicity 100 Myr tracks (blue dotted curve), appropriate for
Pleiades-age objects.

By combining the spectroscopically-constrained parameters \teff\, and
\logg\, alone, and comparing the values to modern stellar evolutionary
models, we infer that the age of $\kappa$ And is almost certainly in
the range $\sim$50-400 Myr. The well-constrained combination of
\teff\, and \logg\, estimated by \citet{Fitzpatrick05} for $\kappa$
And A is consistent with age $\sim$300 Myr for subsolar composition
([M/H] = -0.36) and age $\sim$180 Myr for solar composition. Using the
rotating and non-rotating tracks of \citet{ege12}, one finds the
spectroscopic parameters of \citet{Fitzpatrick05} for $\kappa$ And A
consistent with ages of $\sim$220 Myr and $\sim$200 Myr, respectively.
We conclude that the combination of \teff\, and \logg\, for $\kappa$
And are consistent with an isochronal age of $\sim$200 Myr, however it
may be as old as $\sim$300 Myr if the star is indeed metal poor. As we
show in the next section, these age estimates are commensurate with
that inferred through comparison of the HR diagram position to
evolutionary tracks.

\subsubsection{Luminosity vs. \teff\ Analysis}\label{sec:lum_versus_teff}

\begin{figure}
\epsscale{1.20}
\plotone{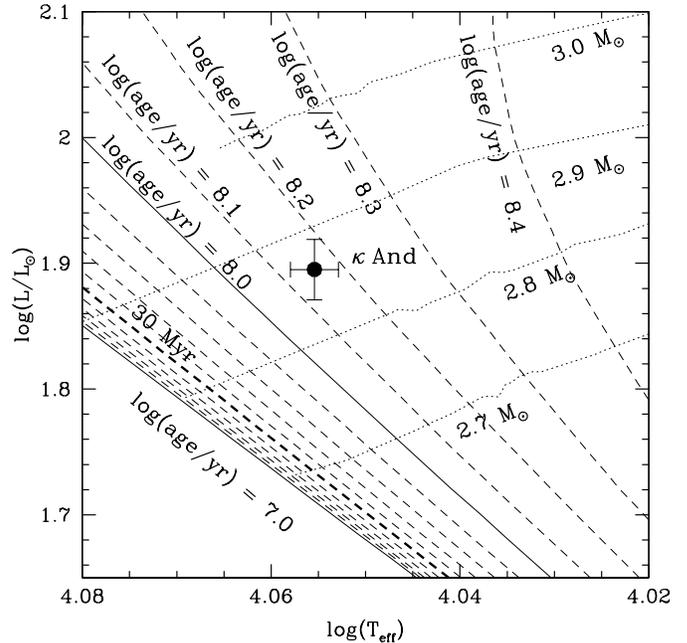}
\caption{Theoretical HR diagram position for $\kappa$ And with
  \citet{Bertelli09} evolutionary tracks for solar composition (Y =
  0.27, Z = 0.017) overlain.  The 30 Myr isochrone (log(age/yr) =7.5)
  is shown as a thick dashed line. Using these tracks, $\kappa$ And
  has age 140 Myr.  Tracks which include rotation and lower
  metallicity produce systematically older ages. Taking into account
  uncertainties in the composition (assuming the star has bulk
  composition ranging from [Fe/H]=-0.36 to solar), we estimate an
  isochronal age of 220\,$\pm$\,100 Myr. }
 \label{fig:HRD}
\end{figure}

In Figure \ref{fig:HRD}, we plot the HR diagram position for $\kappa$
And (adopting the \teff\, from Fitzpatrick \& Massa (2005), with the
revised luminosity from \S~\ref{sec:stellar_params} along with
evolutionary tracks and isochrones from \citet{Bertelli09} assuming
approximately protosolar composition (Y = 0.27, Z = 0.017).
Sampling within the \teff and luminosity uncertainties using Gaussian
deviates, we find that the HR diagram position is consistent with an
age of 140\,$\pm$\,17 Myr and mass 2.89\,$\pm$\,0.03 \msun.
Adopting the \citet{Bertelli09} tracks for a slightly lower (yet
plausible) helium mass fraction (Y = 0.26, Z = 0.017), the HR diagram
point is consistent with age 139\,$\pm$\,17 Myr (2.90\,$\pm$\,0.03
\msun).
If we decrease the metal fraction by $\Delta$Z = 0.001 (Y = 0.27, Z =
0.016), this shifts the age slightly older: 152\,$\pm$\,16 Myr
(2.84\,$\pm$\,0.03 \msun).
Lowering the metal fraction to levels suggested for the proto-Sun
informed by recent observations using 3D solar atmosphere models
\citep[e.g.][]{Asplund09} (Y = 0.27, Z = 0.014), one would derive
177\,$\pm$\,15 Myr (2.79\,$\pm$\,0.03 \msun).
We can also estimate an isochronal age which assumes that the measured
photospheric metallicity is indicative of the star's bulk composition
([Fe/H$\sim$-0.36]).  We scale the star's chemical composition by
assuming a linear trend in $\Delta$Y/$\Delta$Z = 1.57, which connects
the Big Bang primordial abundances \citep[Y = 0.248, Z = 0.00;
][]{Steigman10} with the solar photospheric ratio (X/Z) and protosolar
Y estimated by \citep{Asplund09}.  Adopting the metallicity from
\citet{Fitzpatrick05} ([Fe/H] = -0.36), we interpolate an approximate
chemical composition of Y = 0.26, Z = 0.006. Using this subsolar
chemical composition, we infer that the HR diagram position of
$\kappa$ And would be consistent with age 317\,$\pm$\,10 Myr and mass
2.52\,$\pm$\,0.03 \msun. Note that this chemical composition
represents almost certainly a strong lower limit to the plausible
helium and metal mass fractions, and hence defines an upper limit on
the star's age and a lower limit on its mass.

As a check, we evalulate the HR diagram position of $\kappa$ And A
using other sets of tracks. Using the \citet{Girardi00} evolutionary
tracks for [Fe/H] = 0.0\,$\pm$\,0.1 via the online isochrone
interpolator PARAM
1.1\footnote{\url{http://stev.oapd.inaf.it/cgi-bin/param\_1.1}}, we
find that $\kappa$ And's HR diagram position\footnote{Instead of
  inputting luminosity directly, we entered the V magnitude and
  parallax listed in Table 2, along with the \citet{Fitzpatrick05}
  \teff\, and metallicity.} corresponds to age 252\,$\pm$\,33 Myr and
mass 2.60\,$\pm$\,0.06 \msun, with surface gravity \logg\, =
4.12\,$\pm$\,0.02. Assuming [Fe/H] = 0, the same tracks yield an age
of 121 Myr, mass of 2.85 \msun\, and \logg\, = 4.17.  Using the
rotating evolutionary tracks from \citet{Georgy13} for their assumed
solar composition (Z = 0.014) and $v_{rot}$/$v_{crit}$ = 0.3, $\kappa$
And's age is approximately 250 Myr for mass 2.75 \msun.  Combining our
estimate of the mass of $\kappa$ And A ($\sim$2.8 \msun) with our
updated radius estimate in Table 2 (2.29 \rsun) leads to an estimate
of the star's critical rotational velocity of $\sim$480 \kms, hence
for \vsini\, = 150 \kms \citep{Abt02}, $v_{eq}$/$v_{crit}$ $>$
0.3. Hence, the evolutionary tracks that include rotation which show
slightly older ($\sim$10\%) ages, are probably to be favored.

If the star's bulk composition is similar to solar (Z $\simeq$
0.015-0.017), the age is likely to be $\sim$180\,$\pm$\,70 Myr. If the
star's bulk composition reflects its photospheric abundances (Z
$\simeq$ 0.006), then the star may be of order $\sim$250\,$\pm$\,70
Myr. Hence, there is a systematic uncertainty in the age at the
$\sim$40\%\, level due to the uncertainty in the bulk metal fraction
of the star. Uncertainties due to the helium fraction, observational
uncertainties, rotation, and other differences between the input
physics of the different stellar evolutionary models, each contribute
to the age uncertainty at the $\sim$10\%\, level.  {\it We conclude
  that the HR diagram position for the star is consistent with an
  approximate age of 220\,$\pm$\,100 Myr and mass 2.8$^{+0.1}_{-0.2}$
  \msun}. The derived isochronal age range from the HR diagram
analysis is commensurate with that from the \teff\, vs. \logg\,
analysis in \S3.2.2.

%
%
%

Our new estimate of 220\,$\pm$\,100 Myr is $\sim$7$\times$ older than
the age estimate presented by \citet{Carson13}. Based on the
combination of \teff, \logg, and luminosity, an age for $\kappa$ And A
younger than 120 Myr or older than 320 Myr seem extremely unlikely.
If the bulk composition of $\kappa$ And is truly as metal poor as the
photosphere ([Fe/H] $\simeq$ -0.3), then not only is $\kappa$ And
$\sim$10$\times$ older than the 30 Myr-old Columba association, but
its chemical composition contains less than half the metals of other
Columba members.

With this revised age in hand, we use the DUSTY models of
\citet{cba00} to estimate the mass of $\kappa$ And B. Using the
$L$-band photometry from \citet{bcm13} with our revised age of
220$\pm$100 Myr, we find a revised mass of 50$^{+16}_{-13}$ \mjup,
where the uncertainty is driven almost entirely by our derived
uncertainty in the age of $\kappa$ And A.




\subsection{Multiplicity}\label{sec:multiplicity}

High mass stars show a high degree of multiplicity \citep[e.g.][and
references therein]{dk13} and characterizing the multiplicity of the
$\kappa$ And system, and hence the contributions to the observed
system luminosity, has significant implications for its age
(\S\ref{sec:age}). An equal-flux binary companion would significantly
bias the inferred luminosity (\S\ref{sec:stellar_params}), lowering
the \logl~value in Figure~\ref{fig:HRD} by 0.3 dex, placing it near
the 30 Myr age track.  Hence, understanding the multiplicity of this
system is crucial for a correct interpretation of
Figure~\ref{fig:HRD}. Further, low-mass binary companions could be
useful as additional age indicators. There have been numerous
observations of $\kappa$ And with various techniques, but they have
not been synthesized into a single set of limits. We therefore present
new observations with nonredundant aperture-mask interferometry, as
well as interpreting existing radial velocity and imaging data in the
Appendix, and compile a comprehensive limit on the existence of binary
companions at all semimajor axes (from $10^{-2}$ AU to $10^4$ AU).

\subsubsection{New Limits from Nonredundant Mask Interferometry}


The technique of Aperture Masking Interferometry (sometimes referred
to as ``Sparse Aperture Masking'' or ``Non-Redundant Masking'') is now
well-established as a means of achieving the full diffraction limit of
an AO-equipped telescope \citep[][and references therein]{lmi06,
  kim08, lta11, hci11}.  We obtained new aperture masking observations
of the $\kappa$ And system on 2012 December 2 UT, using Keck II and
the facility adaptive optics imager (NIRC2). To maximize resolution
and sensitivity to short-period binary companions, we used the
$J_{cont}$ filter and an 18-hole aperture mask. For calibration, we
also observed the stars HIP 114456 and HIP 116631.

$\kappa$ And was observed with two sets of 10 individual 10s
integrations, and we observed each calibrator for one such
observation.  The data analysis follows the same prescription as in
\citet{kim08, hci11, kih11}, so we refer the reader to these works. We
summarize the detection limits as a function of projected separation
in Table~\ref{tab:detlims}.

\subsubsection{Limits on Stellar Binary Companions}

\begin{figure}
\epsscale{1.25}
\plotone{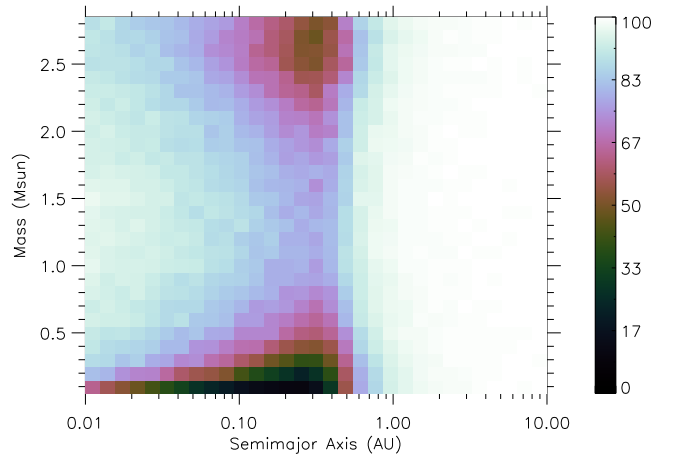}
\caption{The percentage of stellar binary companions, as a function of
  companion mass and semimajor axis, that would have been detected by
  the radial velocity, aperture masking, and direct imaging
  observations that we summarize in Section~\ref{sec:multiplicity} as
  well as this paper's appendix. The vertical color scale bar shows
  this percentage ranging from 0 to 100\%.  Nearly all stellar
  companions with $a \ga 0.6-0.7$ AU are ruled out by the aperture
  masking observations, while radial velocities rule out the majority
  of short-period stellar companions with $M \ga 0.5$ $M_{\odot}$.}
\label{fig:binlimits}
\end{figure}

Utilizing the information on wide binary companions contained in the
appendix, as well as archival radial velocities also listed in the
appendix, we have combined all of the above data in a unified Monte
Carlo simulation that computes detection rates as a function of
companion mass and semimajor axis. We computed $10^3$ randomly
generated orbits across a grid with bins of 0.1 $M_{\odot}$ in $M_{\rm
  secondary}$ (spanning 0.1--2.8 $M_{\odot}$) and 0.1 dex in $\log(a)$
(spanning $10^{-2}$ to $10$ AU); we did not test wider separations
because the aperture masking and Subaru coronagraphic observations
published in \citet{Carson13} rule out all stellar companions, and the
radial velocity data is not useful for substellar companions. For each
randomly generated orbit in the Monte Carlo, we tested the $\chi^2$
goodness of fit for the radial velocity time series while also
verifying that the companion would not have been detected in any of
the direct imaging epochs. We regarded a companion to be ``ruled out''
if the $\chi^2$ statistic is larger than the 95\% confidence limit
(i.e., the orbit would have produced a signal at $>$95\% confidence).
We present the resulting limits on stellar binary companions in
Figure~\ref{fig:binlimits}, which shows the percentage of stellar
binary companions, as a function of companion mass and semimajor axis,
that would have been detected by the radial velocity, aperture
masking, and direct imaging observations.  Nearly all stellar
companions with $a \ga 0.6-0.7$ AU are ruled out by the aperture
masking observations, while radial velocities rule out the majority of
short-period stellar companions with $M \ga 0.5$ $M_{\odot}$.

We note that there exists a region in which nearly equal mass (2-3 \msun) binary companions are not completely ruled out. However, in their work reporting observations with the Palomar Testbed Interferometer (PTI), \citet{vvc08} did not find this system to be resolved. Further, given the resolution and field-of-view of PTI, this work should have reported a similar-brightness binary companion in their data. They claimed that even given the poor fit to a single point source, then the noise in the visibilities would have been consistent with a companion showing ~4 magnitudes of K-band contrast or more, rather than 0 or 1 magnitudes of contrast. Thus, we can appeal to their results to argue that nothing lies in the regime which is not formally ruled out by our analysis.

Our analysis suggests that the luminosity of the $\kappa$ And system is
not biased in any meaningful way by binarity of any kind, and that the
calculated luminosity plotted in Figure~\ref{fig:HRD} is due solely to
a single host star: $\kappa$ And A, reinforcing our isochronal
220\,$\pm$\,100 Myr age estimate.

\begin{deluxetable}{ccll}
\tabletypesize{\scriptsize}
\setlength{\tabcolsep}{0.03in}
\tablewidth{0pt}
\tablecaption{Aperture Masking Interferometry Detection Limits\label{tab:detlims}}
\tablehead{
\multicolumn{2}{c}{Projected Sep} & \colhead{$\Delta J$} & \colhead{$M_{\rm sec}$}\\
\colhead{(mas)} & \colhead{(AU)} & \colhead{(mag)} & \colhead{($M_{Jup}$)}
}
\startdata
10--20 & 0.5--1 & 4.23 & 70\\
20--40 & 1--2   & 5.55 & 32\\
40--80 & 2--4   & 5.46 & 34\\
$>$80  & $>$4   & 5.42 & 35
\enddata
\tablecomments{Limiting companion masses are calculated for the null
  hypothesis that $\tau = 30$ Myr, since that is the hypothesis we are
  trying to disprove. The limiting masses for $\tau = 100$ Myr are
  still $<$0.1 $M_{\odot}$ in all cases.}
\end{deluxetable}


\subsection{Constraints on Inclination}\label{sec:inclination}
In this section we investigate the likelihood that $\kappa$ And A is a nearly pole-on fast-rotator.  Such a configuration could account for the position of $\kappa$ And in the HR diagram (Figure~\ref{fig:HRD}), while still possessing the previously reported age of 30 Myr. 
While effects induced by rapid rotation and inclined viewing angles can lead to scatter in diagrams such as  Figure~\ref{fig:trevor_figure} and color-magnitude diagrams, thereby  confusing the age analysis, such effects are less important for  $\kappa$ And as we show below.  
Specifically, an extreme pole-on orientation is not  possible for $\kappa$~And due to its high observed rotational velocity. However, even very low-inclination models would not change the modelled age noticeably: these models cause the star to become not just more luminous, but also hotter. 
Thus, on an HR diagram, the effect of rotation and inclination is mostly to shift the star along, and not across, an isochrone.

Nonetheless, the observed properties of $\kappa$ And allow us to place some constraints on its inclination. With a projected rotational velocity is $v$sin$i$
= 150 km/s \citep{Abt02}, and taking the approximate mass and radius of
$\kappa$ And listed in Table 3, the formulae of \citet{Townsend04}
predict that the critical rotational velocity for $\kappa$ And to be
394 km/s, which is indeed typical for B9 stars \citep[Table 1
of][]{Townsend04}. Since we do not know the ratio of the star's
equatorial to polar radii, we have adopted the radius inferred from
the luminosity and effective temperature as the star's polar radius in
the formula presented in \citet{Townsend04}.  A lower limit to the ratio of the aspect
ratio $r_{eq}$/$r_{polar}$ can be estimated via the Roche
approximation formula from \citet{Townsend04}, where $v$sin$i$ = 150
km/s places a lower limit on the equatorial velocity of the star.  We
estimate $r_{eq}$/$r_{polar}$ $>$ 1.05. The combination of $v$sin$i$
and predicted critical velocity lead to a constraint on the inclination
of the star: $i$ $>$ 22$^{\circ}$.4. Hence the star can not be within
22$^{\circ}$ of pole-on in orientation.

As \citet{Townsend04} demonstrate, a fast-rotating B star can get a
boost in absolute magnitude and/or reddening in optical color due to
the effects of gravity darkening and viewing angles.  Fig. 3 of
\citet{Townsend04} is instructive for testing whether $\kappa$ And
could be interpreted as a young, extremely fast rotator seen at high
inclination.  In Figure 3 of \citet{Townsend04}, the authors take non-rotating
B-type stars (conveniently including a fiducial B9 dwarf) and
calculate the effects of gravity darkening and viewing angle on $B-V$
color and absolute V magnitude for a range of rotation velocities
(ranging from non-rotating to near critical $v_{eq}$/$v_{crit}$ =
0.95) and at three different inclination angles (0$^{\circ}$,
45$^{\circ}$, 90$^{\circ}$). As discussed previously, the $v$sin$i$
constraints are consistent with $v_{eq}$/$v_{crit}$ $>$ 0.38 and $i$
$>$ 22$^{\circ}$. So we can already rule out $\kappa$ And being a near
face-on star rotating near $v_{crit}$.  For Townsend et al.'s models,
it is the face-on orientation ($i$ = 0$^{\circ}$) that produces the
greatest brightening in absolute magnitude, approximately $\sim$0.6
magnitude in M$_V$ for their most optimistic model ($i$ = 0$^{\circ}$,
$v_{eq}$/$v_{crit}$ = 0.95). The absolute magnitude of $\kappa$ And is
similar to that of late B-type Pleiades ($\sim$120 Myr), and
approximately $\sim$0.4 mag brighter than the ZAMS of
\citet{Schmidt-Kaler82}, which is a reasonable approximation for the
sequence of $\sim$30 Myr late B-type stars. While the $i$ =
45$^{\circ}$ and 90$^{\circ}$ models of \citet{Townsend04} are
plausible for $\kappa$ And, the $i$ = 0$^{\circ}$ is
not. Interpolating amongst the predicted differences in absolute
magnitude and color for the fiducial B9 model of \citet{Townsend04},
it appears that it is extremely difficult to get a plausible model
that can provide a $\sim$0.4 mag boost in absolute magnitude.  For $i$
$\simeq$ 22 (roughly halfway between the $i$ = 0$^{\circ}$ and 45$^{\circ}$
models), $\kappa$ And would have to be rotating at or near critical
velocity i.e. $v_{eq}$/$v_{crit}$ $\simeq$ 0.95. More modest
inclinations of $i$ = 45$^{\circ}$ and 90$^{\circ}$ can not provide
sufficient brightening of the star's real absolute magnitude to the
observed value.

Measurement of a photometric rotation period for the star could provide a measure of the star's rotation, as well as interferometric diameter measurements to test whether the star is consistent with an extreme aspect ratio. 
While included as a suspected variable star in the GCVS catalog
\citep{Samus07}, the Hipparcos survey found the star to be remarkably
photometrically quiet (classified ``C'' = constant), with scatter in Hp
magnitudes of only 0.004 mag, hence measuring a photometric period may
be challenging.

\subsection{Kinematics}\label{sec:kinematics}

Given the evidence presented that the $\kappa$ And system is
$\gtrsim$220 Myr, it is worthwhile to revisit the original
\citet{zrs11} assignment of this system to the 30 Myr Columba
Association.  Using the position, proper motion, and parallax from
\citet{vanLeeuwen07} and mean radial velocity from
\citet{Gontcharov06}, we estimate the velocity of $\kappa$ And to be
($U$, $V$, $W$) = (-11.5\,$\pm$\,0.3, -20.1\,$\pm$\,0.5,
-5.9\,$\pm$\,0.6 ) and ($X$, $Y$, $Z$)= (-16.7, 46.5, -14.8).  As
noted in \citet{Carson13}, the space velocity and position of $\kappa$
And yield $>$95\% probability of Columba membership according to the
moving group prediction method \citep{Malo13}.  However, {\it this 95\%
  probability was only for an additional hypothesis in which a 0.75
  magnitude shift was applied to the photometric sequence for the
  association to account for possible unresolved binarity, a case
  which we rule out in \S\ref{sec:multiplicity}}.  The probability for
a non-binary case carries a lower probability, although this value was
not tabulated in \citet{Malo13}.  Additionally, further investigation
reveals that $\kappa$ And was included as a bona fide member of the
collection of stars comprising the Columba group kinematics in this
work.  This fact automatically increases its derived probability for
membership in Columba.


\begin{figure}
 \epsscale{1.2}
\plotone{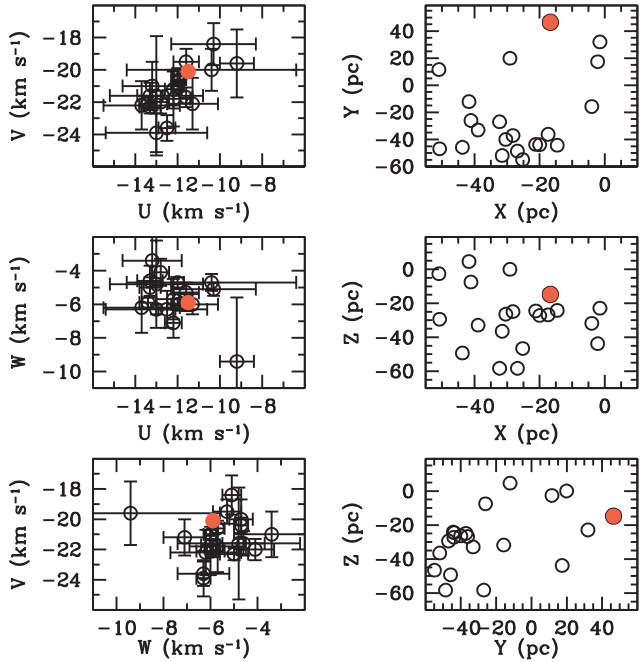}
 \caption{The UVWXYZ velocities and positions for the 20 bona fide
   Columba members (open circles) listed in \citet{Malo13} while the filled (orange) symbol indicates the $\kappa$ And system. While the UVW values for $\kappa$ And
   are in agreement with those for the Columba group, it has the
   largest Y position of the group.  }
 \label{fig:uvwxyz}
 \end{figure}

Figure~\ref{fig:uvwxyz} shows the UVWXYZ velocities and positions for
the 21 bona fide Columba members listed in \citet{Malo13} with the
$\kappa$ And system highlighted.  Figure~\ref{fig:uvwxyz} shows that
the $UVW$ velocities for $\kappa$ And are consistent with the 20 other
bona fide members of Columba.  However it is a 2.7$\sigma$ outlier in galactic Y
position, having the largest Y position of the entire Columba ensemble.  Even further, \citet{zrs11} do not use the galactic Y position as part of their criteria for
moving group membership.  



The strong agreement between the $UVW$ velocity of $\kappa$ And B and the Columba association make a full kinematic ``traceback'' analysis challenging.  Despite the fact that this object is a significant positional outlier in the $Y$-direction, any discussion on the star's past position, velocity relative to the centroid of the Columba association, etc. is weakened due to the similarities in velocities of the star and the Columba group discussed above.  Further, the errors in velocities for $\kappa$ And A and Columba are sizeable ($\sim$0.5 km/s and $\sim$1 km/s, respectively).  Nonetheless, adopting the centroid position and $UVW$ velocity for Columba listed in \cite{Malo13} , and the $UVW$ for $\kappa$ And A listed in Table 1, the star is currently ~80 pc away from the centroid of Columba, and its velocity differs by ~1.5 km/s.  Using an epicycle orbit approximation code, it appears that $\kappa$ And was only slightly closer to the Columba centroid in the past: ~18 Myr ago it was ~60 pc from Columba, and ~30 Myr ago it was ~74 pc from Columba.

However, the outlying $Y$ position of $\kappa$ And (46.5pc) raises questions about the likelihood of its formation near the Columba groups centroid ($Y$=-31.3). Notably, for $\kappa$ And to have formed near Columba's centroid 30 Myr ago, it would have had to inherited a peculiar V velocity of $\Delta V$ = (46.5 + 31.3pc)/(30 Myr) = 2.59 pc/Myr $\sim$ 2.6 km/s.  Given Columba's current $V$ velocity of $V$=-21.3, a ``runaway'' star would have velocity $V+\Delta V \simeq$ -18.7 km/s. However, this is still within $\sim$1$\sigma$ of what is observed for $\kappa$ And.


\section{Summary}\label{sec:summary}

In this work we have presented analysis of the spectra and photometry
of the companion $\kappa$ And B, as well as presented a comprehensive
analysis of the age, multiplicity, and moving group kinematics of the $\kappa$ And AB
system.  We summarize our results as follows:

\begin{itemize}
\item $YJH$-band low resolution spectra obtained through high
  contrast imaging with Project 1640 are consistent with an
  intermediate age ($\lesssim$300 Myr) brown dwarf with L1$\pm$1 spectral type, although similiarities with field mid-L objects are present.
 
\item By fitting synthetic models to the Project 1640
  spectrophotometry, we constrain the surface gravity and effective
  temperature of $\kappa$ And B to be \logg=4.33$^{+0.88}_{-0.79}$ and  \teff=2040K$^{+58}_{-64}$, respectively.

\item Comparing these photospheric properties to theoretical isochrones in \logg~and
  \teff~parameter space indicates an age much older than 
  the 30 Myr age reported previously for $\kappa$ And B. 

\item Previously published \logg\ and \teff\ values for $\kappa$ And A
  are compared to theoretical isochrones, indicating
  ages of $\sim$100-300 Myr. The HR diagram position of $\kappa$ And A
  is consistent with the same age for a range of assumed chemical
  compositions.  Taken together, the stellar parameters are consistent
  with an isochronal age of 220\,$\pm$\,100 Myr, where the age uncertainty
  is dominated by the star's chemical composition.

\item We combine aperture masking interferometry, archival radial
  velocity data from the literature, and archival multi-epoch imaging
  of $\kappa$ And A to rule out any faint stellar companions beyond
  $\sim$0.6 AU (Figures~\ref{fig:binlimits} and \ref{fig:ALKfigure})
  that could be causing the star to be overluminous for the
  originally-quoted 30 Myr age. In addition, we show that a nearly ``pole-on'' viewing angle coupled with extremely rapid rotation is unlikely to be the configuration contributing to this star's overluminosity.  

\item $\kappa$ And A appears to be a kinematic outlier compared to other
  Columba members. 
  While the velocity of $\kappa$ And is consistent  with that of other Columba members, its Galactic Y position is an outlier.  Taken together with its overluminosity and low surface gravity expected for a 30 Myr old, late-B star, $\kappa$ And is most likely an interloper to the Columba association.


\item Through the use of Hertzsprung-Russell diagram analysis as well as
comparison of the \logg\ and \teff\ parameters for $\kappa$ And A with
theoretical isochrones, we have shown that the star has an 
age closer to 220 Myr than the originally assumed 30 Myr based on association with Columba. 
These ages indicate that the mass of $\kappa$ And B is 50$^{+16}_{-13}$ \mjup, rather than the previously claimed 12-14 Jupiter Masses.


\end{itemize}

\acknowledgements
We thank the anonymous referee for numerous helpful suggestions. 
SH is supported by an NSF Astronomy and Astrophysics Postdoctoral
Fellowship under award AST-1203023.  LP performed this work in part under contract
with the California Institute of Technology funded by NASA through the
Sagan Fellowship Program.  ALK was supported by a Clay
Fellowship.  EEM is supported by NSF award AST-1008908 and generous
donations of Gabriela Mistral Pisco.  GV is supported by a NASA OSS
grant NMO7110830/ 102190.   RN performed this work with funding through
a grant from Helge Axson Johnson's foundation.  JRC is supported by
NASA Origins of Solar Systems Grant NNX13AB03G.  A portion of this
work is or was supported by the National Science Foundation under
Grant Numbers AST-0215793, 0334916, 0520822, 0804417 and 1245018.  A
portion of the research in this paper was carried out at the Jet
Propulsion Laboratory, California Institute of Technology, under a
contract with the National Aeronautics and Space Administration and
was funded by internal Research and Technology Development funds.  Our
team is also grateful to the Plymouth Hill Foundation, and an
anonymous donor, as well as the efforts of Mike Werner, Paul Goldsmith
and Jacob van Zyl.  Any opinions, findings, and conclusions or
recommendations expressed in this material are those of the authors
and do not necessarily reflect the views of the National Science
Foundation.  Finally, the entire team expresses is sincere gratitude
and appreciation for the hard work of the Palomar mountain crew,
especially by Steve Kunsman, Mike Doyle, Greg van Idsinga, Bruce
Baker, Jean Mueller, Kajsa Peffer, Kevin Rykowski, Carolyn Heffner and
Dan McKenna.  This project would be impossible without the
flexibility, responsiveness and dedication of such an effective and
motivated staff.

\appendix

\section{Limits on Wide Companions to $\kappa$ And}

Here we present limits on wide binary companions to the $\kappa$ And
system as well as a listing of archival radial velocities. These
results are incorporated into our analysis presented in
\S\ref{sec:multiplicity}.


 \begin{figure*}
 \epsscale{1.1}
\plottwo{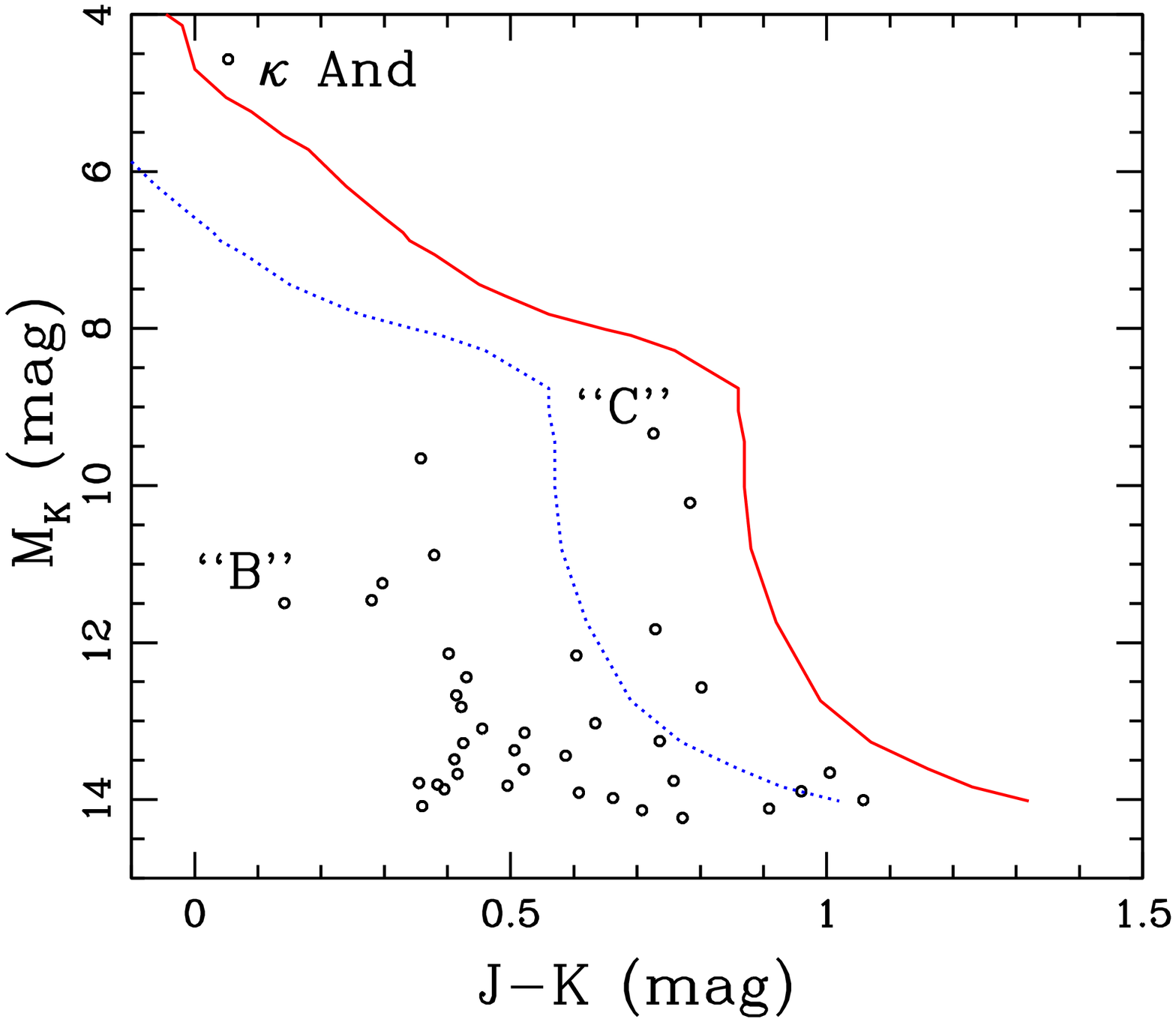}{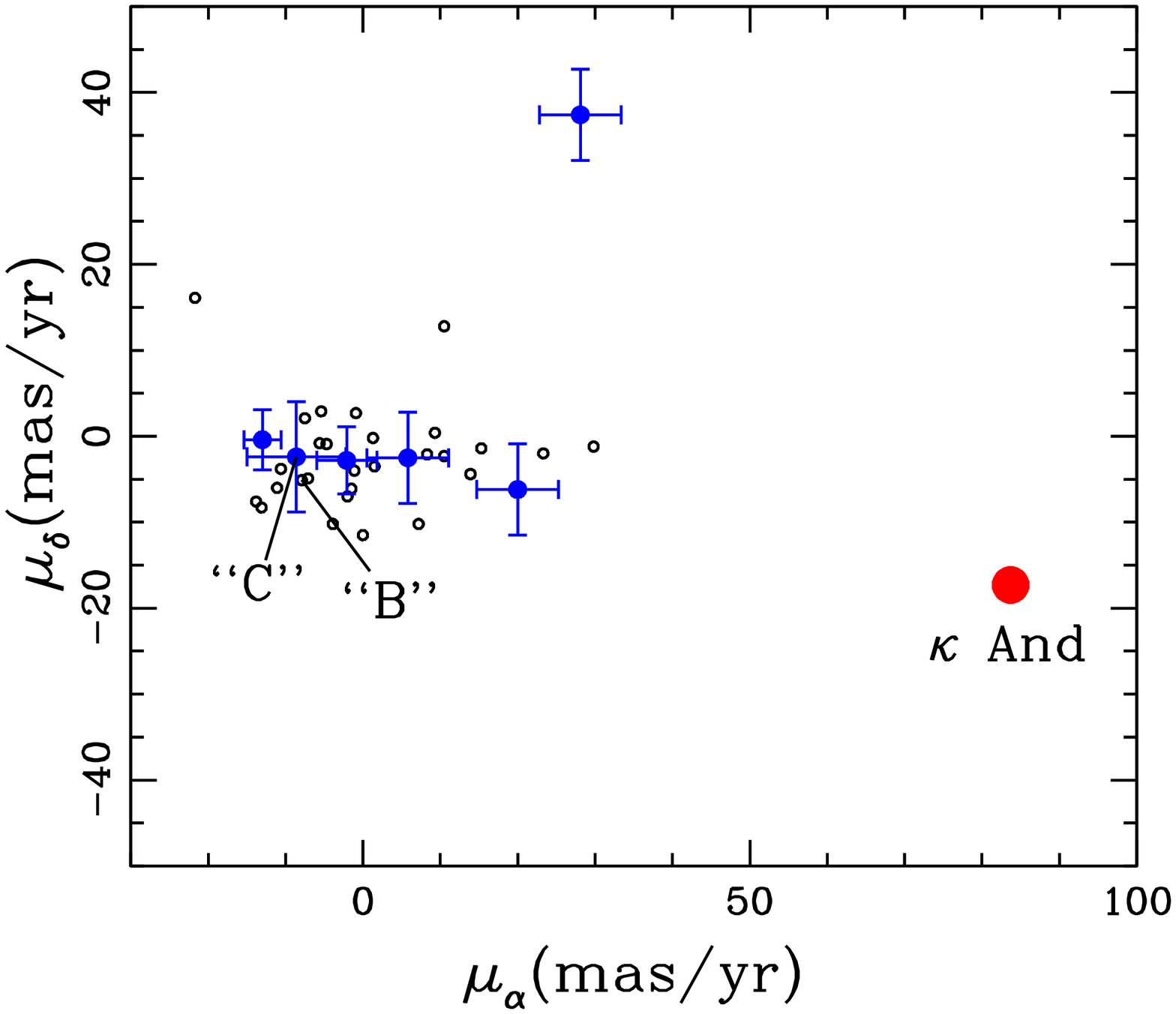}
 \caption{Left: A 2MASS ($M_K$,$J-K$) color-magnitude diagram for the
   38 sources with $K < 14.3$ that are located within $\rho =
   $20--300\arcsec\, of $\kappa$ And. The solid red line is the main
   sequence at the distance of $\kappa$ And \citep{kh07}, while the
   blue dotted line shows the $\Delta(J-K)<0.3$ limit which denotes
   possible consistency. Only seven sources (including ``C'', but
   excluding ``B'') have colors which are marginally consistent with
   physical association. Right: A proper-motion diagram for the 34
   sources which have catalog proper motions. The blue points denote 6
   of the 7 sources with marginally consistent colors. None of these
   sources (including both ``B'' and ``C'') are comoving with $\kappa$
   And. Visual inspection of the original POSS-I red plate (epoch
   1952) and the 2MASS $K$ band image (epoch 1998) show that the
   remaining four sources have proper motions of $<$20 mas/yr, and
   hence also are not comoving. We therefore conclude that the
   purported ``B'' and ``C'' components are not physically associated,
   and neither are any other sources with $M_{lim} > 15$ $M_{Jup}$ (if
   $\tau = 30$ Myr) and $\rho = $1000-15000 AU.}
 \label{fig:ALKfigure}
 \end{figure*}

\subsection{Literature Observations}

The primary star $\kappa$ And A has been observed by numerous radial
velocity surveys over the past century; we list those radial velocity
observations which we could recover in
Table~\ref{tab:rvs}. \citet{pwj68} reported 11 radial velocity
measurements for $\kappa$ And that were taken between 1960 October 4
and 1961 November 4 (UT), and found a mean radial velocity of $v =
-15$ km/s with a standard deviation of $\sigma = 10.5$ km/s and a
standard deviation of the mean of $\sigma_{\mu} = 3$ km/s. \citet{h37}
reported 3 radial velocity measurements that were taken between 1923
Sep 13 and 1926 Dec 14 (UT), and found a mean radial velocity of $v =
-19$ km/s with a standard deviation of $\sigma = 5$ km/s and a
standard deviation of the mean of $\sigma_{\mu} = 3$ km/s; since there
are only three epochs and we must be concerned with zero point shifts,
we do not use these data. \citet{w53} reported that $\kappa$ And had a
mean radial velocity of $v = -9.0$ km/s for 10 observations, but did
not report an uncertainty or the individual measurements, so we cannot
use these measurements either. Finally, we also note that \citet{hg08}
reported that $\kappa$ And is a relatively fast rotator ($v_{rot} =
169$ km/s), and therefore even if it is a low-amplitude double-lined
spectroscopic binary, radial velocity measurements might not resolve
the individual components. Our analysis therefore must consider the
shift in spectral line centroids in placing constraints on the
presence of a double-lined spectroscopic binary.

\subsection{Limits on Wide Co-moving Companions with Archival Multi-Epoch Imaging}

Even before the discovery of $\kappa$ And ``b'', $\kappa$ And was
considered a binary (ADS 16916, WDS 23404+4420, HJ 1898).
J.F.W. Herschel reported \citep{h31} possible companions to $\kappa$
And at $\rho = 35$\arcsec\, (``B'') in epoch 1828 and $\rho =
98$\arcsec\, (``C'') in epoch 1836 \citep{s44, Mason01}. The nearest
2MASS counterparts for these stars are 2MASS J23402285+4419177 ($\rho
= 48$\arcsec) and 2MASS J23401480+4420469 ($\rho = 113$\arcsec). If
these (or any other) stars were indeed associated, then they would
offer a valuable check on the age of the system. To test the
association of these candidates and to search for other potential
comoving companions, we have investigated the nature of all
identifiable sources within $<$5\arcmin\, ($<$15000 AU) of $\kappa$
And.

We queried the 2MASS Point Source Catalog (which has the highest image
fidelity) to identify 38 candidate companions with $K < 14.3$ and
$\rho < 5$\arcmin. The PSC clearly detected a source with $K=13.9$ at
$\rho = 27$\arcsec, and the background flux is similar down to $\rho =
20$\arcsec. We therefore estimate that any source brighter than the
2MASS detection limit ($K=14.3$ at 10$\sigma$) would have been
detected at $\rho > 20$\arcsec. We also compiled proper motions for
most of these sources from UCAC4 \citep[for 9 sources;][]{Zacharias12}
and from PPMXL \citep[for 25;][]{Roeser10}. Four candidate companions
did not have proper motions in either catalog, but in all cases,
visual inspection of the raw images showed that they moved by
$<$1\arcsec\, ($\la$20 mas/yr) between the POSS-I epoch (1952) and the
2MASS epoch (1998).

In Figure ~\ref{fig:ALKfigure} (left), we show a ($J-K$,$K$)
color-magnitude diagram for the 38 sources with $K < 14.3$ identified
by 2MASS. Figure ~\ref{fig:ALKfigure} also shows the main sequence for
stars located at the distance of $\kappa$ And. Only 7 sources are
located within $\Delta(J-K) < 0.3$ mag of the main sequence; the
remaining 31 sources (including the ``B'' companion from 1831, as well
as three of the four sources with visually-estimated proper motions)
appear to be unassociated background stars. In Figure
~\ref{fig:ALKfigure} (right), we show a proper motion diagram for the
34 sources with measured proper motions. None agree with the HIPPARCOS
proper motion for $\kappa$ And to within 3$\sigma$ (including both
``B'' and ``C'' objects from 1831), and hence all (including the
fourth star with a visually estimated proper motion limit) appear to
be unassociated background sources. We therefore conclude that there
are no comoving companions (including the purported ``B'' and ``C''
companions) with $K < 14.3$ ($M_{lim} > 15$ $M_{Jup}$, for the null
hypothesis of $\tau = 30$ Myr) located within $\rho =$20--300\arcsec\,
($\rho =$ 1000--15000 AU) of $\kappa$ And.



\begin{deluxetable}{ccll}
\tabletypesize{\scriptsize}
\setlength{\tabcolsep}{0.03in}
\tablewidth{0pt}
\tablecaption{Radial Velocities\label{tab:rvs}}
\tablehead{
\colhead{Epoch} & \colhead{$v$} & \colhead{$\sigma_v$} & \colhead{Source}
\\
\colhead{(JD)} & \colhead{(km/s)} & \colhead{(km/s)}
}
\startdata
2437212.34 &  -9 & 7 & \citet{pwj68}\\
2437215.41 & -14 & 3 & \citet{pwj68}\\
2437222.44 & -14 & 5 & \citet{pwj68}\\
2437223.36 &  +9 & 9 & \citet{pwj68}\\
2437230.50 &  -9 & 8 & \citet{pwj68}\\
2437243.45 & -29 &10 & \citet{pwj68}\\
2437558.50 & -25 & 4 & \citet{pwj68}\\
2437587.41 &  -4 & 8 & \citet{pwj68}\\
2437590.46 & -21 & 3 & \citet{pwj68}\\
2437601.52 & -17 & 5 & \citet{pwj68}\\
2437608.38 & -18 & 7 & \citet{pwj68}\\
2423676.791 & -19.4 & ... & \citet{h37}\\
2423676.818 & -13.4 & ... & \citet{h37}\\
2424963.659 & -24.2 & ... & \citet{h37}\\
\enddata
\end{deluxetable}

\end{document}